# Is Quintessence an Indication of a Time-Varying Gravitational Constant?

By:  Dr. Christopher Pilot
     Department of Physics
     Gonzaga University
     Spokane, WA 99202

     pilot@gonzaga.edu

**ABSTRACT:**

A model is presented where the quintessence parameter, w, is related to a time-varying gravitational constant. Assuming a present value of w = -.98, we predict a current variation of $\dot{G}/G$ = - .06 $H_0$, a value within current observational bounds. $H_0$ is Hubble's parameter, G is Newton's constant and $\dot{G}$ is the derivative of G with respect to time. Thus, G has a cosmic origin, is decreasing with respect to cosmological time, and is proportional to $H_0$, as originally proposed by the Dirac-Jordan hypothesis, albeit at a much slower rate. Within our model, we can explain the cosmological constant fine-tuning problem, the discrepancy between the present very weak value of the cosmological constant, and the much greater vacuum energy found in earlier epochs (we assume a connection exists). To formalize and solidify our model, we give two distinct parametrizations of G with respect to "a", the cosmic scale parameter. We treat $G^{-1}$ as an order parameter, which vanishes at high energies; at low temperatures, it reaches a saturation value, a value we are close to today. Our first parametrization for $G^{-1}$ is motivated by a charging capacitor; the second treats $G^{-1}(a)$ by analogy to a magnetic response, i.e., as a Langevin function. Both parametrizations, even though very distinct, give a remarkably similar tracking behavior for w(a), but not of the conventional form, w(a) = $w_0$ + $w_a$ (1-a), which can be thought of as only holding over a limited range in "a". Interestingly, both parametrizations indicate the onset of G formation at a temperature of approximately 7 *$10^{21}$ degrees Kelvin, in contrast to the ΛCDM model where G is taken as a constant all the way back to the Planck temperature, 1.42 * $10^{32}$ degrees Kelvin. At the temperature of formation, we find that G has increased to roughly 4*$10^{20}$ times its present value. For most of cosmic evolution, however, our variable G model gives results similar to the predictions of the ΛCDM model, except in the very early universe, as we shall demonstrate. In fact, in the limit where w approaches -1, $\dot{G}/G$ vanishes, and we are left with the concordance model. Within our framework, the emergence of dark energy over matter at a scale of a ≈ .5 is that point where $G^{-1}$ increases noticeably to its current value $G_0^{-1}$. This weakening of G to its current value $G_0$ is speculated as the true cause for the observed unanticipated acceleration of the universe.

# I  INTRODUCTION

Quintessence is a hypothetical form of dark energy based on a dynamical scalar field whose value changes with respect to cosmological time. Its equation of state relates the pressure of the vacuum to its density, and this equation is determined by the potential energy term as well as a kinetic term involving the scalar field. This is to be contrasted with the concordance ΛCDM model where we have a cosmological constant, which does not scale. In fact, there the quintessence parameter, w, relating pressure to density is by definition precisely equal to -1, indicating that the pressure and density are fixed, where pressure is always equal to the density but negative. While the ΛCDM model is highly successful, quintessence is studied because a) it may help us better understand the true nature of dark energy (the ΛCDM model provides no explanation of either dark energy or dark matter), b) it could help us resolve the cosmological constant fine-tuning problem, and c) it may help us understand the coincidence problem, which seeks to address the question as to why now for the unanticipated acceleration of the universe. Why is the vacuum density parameter, $\Omega_\Lambda$, comparable to the matter density parameter, $\Omega_{MATTER,}$ in the present epoch? If the universe had accelerated at an earlier era due to dark energy, then we would not have the structure we see today.

The cosmological fine tuning problem [1] is a vast discrepancy between the present, observed value for the cosmological constant, $\Lambda_{OBS.} = 1.11 * 10^{-52}$ m$^{-2}$ = $4.33 * 10^{-66}$ (eV)$^2$ in natural units, and the vacuum energy at the Planck scale, $\Lambda_{VACUUM}$ = (Planck Length)$^{-2}$ = $3.83 * 10^{69}$ m$^{-2}$. The latter amounts to $(1.22 * 10^{28}$ eV$)^2$ in natural units. The cosmological constant and the vacuum are often identified with each other in cosmology, i.e., the cosmological constant is assumed to be a characteristic of the vacuum. We also assume such a connection exists. Numerically, the discrepancy between the cosmological constant and the vacuum energy is $(1.22 * 10^{28}$ eV$)^2$/ $4.33 * 10^{-66}$ (eV)$^2$ = $.345 * 10^{122}$, often referred to as the worst fine-tuning problem in physics. There are other reasons for considering quintessence, but these are among the major ones cited. For several good reviews on the subject, we refer the reader to references [2,3]. Some original articles on quintessence are found in references [4-10].

Perhaps another reason for considering quintessence, is the observational fact that the quintessence parameter, w, does not appear to equal precisely negative one as would be required in the concordance model. This number is difficult to determine, and yet over the past decade, its value seems to center about [11-14] -.97 to -.98, if one assumes a flat universe. Is this an indication of physics beyond the ΛCDM model, as some researchers suspect? Within observational uncertainty, however, it should be noted that the uncertainty in "w" does include the w = -1 value required for ΛCDM.

The observational limits set on "w" depend on the tests employed. The most stringent limit on "w" at present uses high z supernovae data and assumes that space is flat. Combined with WMAP and BAO (Baryon Acoustic Oscillation) data, that limit is established as [15,16]

$$w = -.98 +/- .053 \qquad (68\% \text{ CL}) \qquad \text{WMAP + BAO + SN} \qquad (1-1)$$

The flat space assumption, where the density parameter $\Omega_k = 0$, provides good constraints on "w"; $\Omega_k \neq 0$, on the other hand, provides poor limits on "w" unless $\Omega_k$ is also specified. If space is not assumed flat, and if we select a particular value for $\Omega_k$, then the following limits are obtained [10]

$$\Omega_k = -.0057 \, ^{+.0060}_{-.0062} \qquad (68\% \text{ CL}) \qquad (1\text{-}2a)$$

$$w = -.999 \, ^{+.057}_{-.056} \qquad (68\% \text{ CL}) \qquad (1\text{-}2b)$$

Only when taken together are the limits given in equations (1-2) "tight". In our paper, we shall assume a flat space. On its own, within the assumptions of the ΛCDM model [18], $\Omega_k = -.005^{+.016}_{-.017}$, which is a value so close to zero as to suggest that space is indeed flat. Thus, we will use the result in equation (1-1) as our working ansatz. However, we leave open the possibility that the numerical value of "w" may have to be adjusted in the future.

In this paper, we seek to provide an explanation for quintessence. We argue that it is a manifestation of a cosmic varying gravitational constant, i.e., $G = G(a)$ where "a" is the cosmic scale parameter. Alternatively, $G = G(T)$, where T equals the CMB temperature. Both parametrizations are equivalent due to the identity $T/T_0 = a^{-1} = (1+z)$ where $T_0$ is the present day CMB value equal to 2.725 K, z is the redshift, and "a" is taken to equal unity in the present epoch. A cosmological time varying gravitational constant is an old idea, going back to the works of Paul Dirac and Pascual Jordan. In 1937, P. Dirac [19-21], in his large number hypothesis (LNH), suggested that G is proportional to $t^{-1}$ where "t" equals cosmic time. Almost immediately thereafter, P. Jordan [22-25], embracing this idea as very significant, proposed that $\dot{G}/G = -H$ where H is Hubble's parameter and $\dot{G} = dG/dt$. He also recognized that if G varies, then it must effectively be replaced by a scalar field as a consequence, and he developed such a field theory for a varying G. Both concluded that G decreases with an expanding cosmos. Since the present value of Hubble's parameter [26] is close to 67.74 km/(s Mpc) = $6.925 * 10^{-11}$ yr$^{-1}$, the relative rate of change of G with respect to cosmological time in the present epoch is to be considered not large.

On the observational front, there have been many searches for a non-zero result for $\dot{G}/G$, going back to the mid 1960's. Actually one can go back further. Researchers, already in the 1940's, and 1950's, looked for geological and paleontological evidence for a time-varying G, as for example, in Jordan's expanding Earth hypothesis [27,28]. Trying to establish a geological or paleontological signature for a variable G proved difficult because of the very many complicating factors involved when dealing with the Earth's evolution. Therefore, physicists around the mid 60's started to look for astronomical evidence. Among the earliest astronomical observations, which led to a non-zero signature, was a determination made by Mueller [29], using radar ranging within the solar system from 1966 to 1975. His estimate was $\dot{G}/G = -(3.9 - 9.9) * 10^{-11}$ yr$^{-1}$. Shapiro [30] claimed, more conservatively, that $|\dot{G}/G| < 4 * 10^{-10}$ yr$^{-1}$. Another very early result was obtained between 1981 and 1984 by Van Flandern [31,32], using an entirely different approach. He analyzed lunar mean motion around the Earth, and related the period of orbit with a varying G. Using this method he obtained a value equal to $\dot{G}/G = -(4.5 - 9.3) * 10^{-11}$ yr$^{-1}$. This is very close to the former non-zero estimate of Mueller, and quantitatively in agreement with the Jordan hypothesis. It is to be noted that at the time of these tests, a precise value for Hubble's parameter was not well known. An acceptable value for $H_0$ at that time was debated to lie anywhere between 50 to 100 km/(s Mpc).

More recent observations [33-36] suggest that these values for $\dot{G}/G$ are too high. The limits for $\dot{G}/G$ have been pushed down to about .1 $H_0$ or less, and furthermore, they are more conservative in that most of the recent tests *do not rule out* a zero result. One interesting test analyzes the decay in orbits of binary star systems [37-39] (Nordvedt). He finds that $\dot{G}/G = -(.9 +/- 1.8) * 10^{-11}$

yr$^{-1}$. Another variation test is due to Thorsett [40], who has analyzed the energy release in supernova SNIa explosions, both near (low z) and far (larger z). The look-back times were between 1 Gyr to 12-13 Gyr. Given what is known about SN events, the energy release is proportional to the Chandrasekhar mass, $M_{Ch}$, which in turn is proportional to $G^{-3/2}$. They obtain $\dot{G}/G = - (.6 +/- 4.2) * 10^{-12}$ yr$^{-1}$. In this regard, it was also later recognized [41-43] that rise times for supernova events could be modeled by an analytical formula where the width of the peak of the light curve is given by τ proportional to $M_{Ch}^{1/2}$, which in turn is proportional to $G^{-3/4}$. Thus, distant, i.e., larger z, SN events have supposedly not only smaller peak luminosities, but at the same time, smaller rise times. Rise times between 17.50 +/- .40 and 19.98 +/- .15 days were observed for the far (high z) and near (low z) SN events, respectively. This is presented as further evidence in support of a stronger G value at the time of emission, at earlier look-back times.

Two good up-to-date reviews of the latest observational status on G can be found in references [44,45]. We remark that all the above tests give consistent values for $\dot{G}/G$, of the order of Hubble's value, in spite of the fact that they are obtained using very different methodologies and observations. They also span a period of seven decades of research.

The outline of this paper is as follows. In section II, we make a simple observation and identify G mathematically with "w". A general result is derived, namely, that $\dot{G}/G = -.06\ H_0$ in the present epoch *assuming* we use w = -.98 as is indicated in equation (1-1). In section III, two simple one-dimensional parametrizations for G(a) are presented. Both have the correct limits for an order parameter, which depends on temperature; at high energies (temperatures), the values for $G^{-1}$ vanish and at low temperatures, they assume constant saturation values. We will fix the parameters of both models such that we have a well-defined behavior for $G^{-1}(a)$ in both instances. In section IV, we establish a time-line for $G^{-1}$. It is important to show that the results of our extended models do not deviate too drastically from the well-established ΛCDM model, except in the very early universe. Even though our two parametrizations are quite different, they predict essentially the same features, both qualitatively and quantitatively. In section V, we consider the onset of G formation, i.e., $G^{-1}(a_C)$, where "$a_C$" is the scale parameter at formation. We present arguments for why we believe it occurred at a scale when the CMB temperature was about $7 * 10^{21}$ K. This scale is practically identical in both parametrizations, even though the models are quite distinct from each another, leading us to believe that there may be some universality involved. In the ΛCDM model, G is, of course, a constant up to and including the Planck scale, which is much higher in temperature, $1.42 * 10^{32}$ K. We relax this assumption. Therefore, in the very early universe, our model suggests that cosmic expansion is not hampered or hindered by gravitation; at least not in the form, we currently know it. Finally, in section VI, we present our summary and conclusions.

## II  *A SIMPLE OBSERVATION*

We start with the second version of the Friedmann equations. Written in the present epoch, and at any other cosmological time, we have

$$H_0^2 = 8\pi G_0 \, (\rho_{CRIT0})/3 = 8\pi G_0/3 \, (\Omega_{RAD} + \Omega_{MATTER} + \Omega_\Lambda + \Omega_k) \, (\rho_{CRIT\,0}) \quad (2\text{-}1a)$$

$$H^2 = 8\pi G \, (\rho_{CRIT})/3 = 8\pi G/3 \, (\Omega_{RAD}\, a^{-4} + \Omega_{MATTER}\, a^{-3} + \Omega_\Lambda + \Omega_k\, a^{-2}) \, (\rho_{CRIT\,0}) \quad (2\text{-}1b)$$

All subscripts "0" in this paper refer to the current cosmological epoch. In the above equations, H stands for Hubble's parameter, defined as $H \equiv a^{-1}\,(da/dt) = \dot{a}/a$, where "a" is the cosmological scale parameter. We use the convention where "a" is defined to equal 1 in the present epoch. Going backwards in time, we set a < 1; going forward, we allow for a > 1. The relation, $(\rho CRIT) \equiv (3H^2/8\pi G)$, defines the critical matter/energy density. The cosmic density parameters ($\Omega_{RAD}$, $\Omega_{MATTER}$, $\Omega_\Lambda$, $\Omega_k$) are those associated with radiation, matter (including dark), vacuum and curvature, respectively. By definition, their sum equals one. Unless otherwise stated, current best value estimates for all parameters are obtained using the latest 2015 XIII Planck Cosmological Parameters report [46]. We have $H_0$ = 67.74 +/- .46 km/(s Mpc) = (2.195 +/- .015) * $10^{-18}$ $s^{-1}$. For the critical mass density in the present era, the calculated value is $(\rho_{CRIT0})$ = 8.624 * $10^{-27}$ kg/$m^3$. In addition, for the density parameters, we take ($\Omega_{RAD}$, $\Omega_{MATTER}$, $\Omega_\Lambda$, $\Omega_k$) = (8.3*$10^{-5}$, .3089, .6911, <.005), which conforms to the $\Lambda$CDM model. Finally, in the above, we have $G_0$, which equals Newton's gravitational constant. For what we have in mind we will not automatically assume that $G = G_0 = 6.67 * 10^{-11}$ N*$m^2$/$kg^2$, but rather that G = G(a).

Current evidence suggests that the universe as a whole is remarkably flat, i.e., there is no inherent spatial curvature. Best estimates for $\Omega_k = 1 - \Sigma\,\Omega_i$ suggest that it is less than .005 as shown by the latest Planck data collaboration. Thus, we will also assume that space is flat. Our results would change for a non-flat universe as the parameter "w" would also change. If $\Omega k < 0$, then $\Sigma\,\Omega_i = \Omega_{RAD} + \Omega_{MATTER} + \Omega_\Lambda > 1$ would correspond to a closed universe with positive curvature and $\rho > \rho_{CRIT}$. If $\Omega_k > 0$, then $\Sigma\,\Omega_i < 1$, and this would correspond to an open, i.e. hypergeometric, universe with negative curvature and $\rho < \rho_{CRIT}$.

We will first assume that $G = G_0$. We divide equation (2-1b) by equation (2-1a) to give

$$H^2/H_0^2 = [\Omega_{RAD}\, a^{-4} + \Omega_{MATTER}\, a^{-3} + \Omega_\Lambda\, a^{-3(1+w)}] \quad (2\text{-}2)$$

We have introduced the quintessence parameter, "w" in the last term on the right hand side of equation (2-2). The equation of state for quintessence is $w \equiv p_\Lambda / (\rho_\Lambda\, c^2)$, where $p_\Lambda$ and $\rho_\Lambda$ are the pressure and the mass density associated with the dark energy vacuum, respectively. The parameter "w" can be defined in terms of a scalar field, which we will not go into; for our purposes "w" is a value between zero and -1, equaling the latter in the limit of the $\Lambda$CDM model. If w is set equal to negative one, it is clear from equation (2-2) that dark energy does not scale. If we choose $w \neq -1$, then we allow for scaling. The negative sign for "w" tells us that dark energy is characterized by negative pressure given a positive dark energy density. A current best-fit estimate for "w" is specified by equation (1-1), but only in the limit of a flat space cosmology, which is assumed here.

We next consider the possibility that $G \neq G_0$. We divide equation (2-1b) by equation (2-1a), left hand side by left hand side, and right hand side by right hand side. This renders

$$H^2/H_0^2 = G/G_0\, (\Omega_{RAD}\, a^{-4} + \Omega_{MATTER}\, a^{-3} + \Omega_\Lambda) \quad (2\text{-}3)$$

We next bring in the $G/G_0$ term within the brackets of equation (2-3) as this allows for a comparison with equation (2-2). Upon comparing, we make the identification

$$G/G_0 = a^{-3(1+w)} \equiv a^{-\alpha} \tag{2-4}$$

We have defined $\alpha$ as $\alpha \equiv 3(1-w)$ in equation (2-4). Another way to write equation (2-4) is

$$G/G_0 = \rho_\Lambda / \rho_{\Lambda 0} \tag{2-5}$$

Equation (2-5) follows since $\rho_\Lambda = a^{-\alpha} \, \Omega_\Lambda \, \rho_{CRIT0}$ and $\rho_{\Lambda 0} = \Omega_\Lambda \, \rho_{CRIT0}$. In practice, both G and $\rho_\Lambda$ will vary very slowly for most of cosmic evolution.

A present best estimate for "w" in the current epoch gives a value close to $w_0 = -.98$. We can substitute this value into equation (2-4) to give

$$(G/G_0)|_0 = a^{-.06} \tag{2-6}$$

For $a = 1$, this is a trivial identity. As we shall see shortly, "w" does not increase or decrease appreciably with respect to either temperature or cosmic time if we are close to the present epoch. In fact, "w" is constant for a cosmic scale in the range from .6 to 1.4. Thus equation (2-6) is a very good approximation for G for "a" not too far from $a = 1$, the current epoch. Using equation (2-6), therefore, we can estimate that for the given values, respectively,

$$G/G_0 = (1.013, 1.006, .994, .989) \quad \text{for} \quad a = (.8, .9, 1.1, 1.2) \tag{2-7}$$

These are small deviations about $G_0$. As mentioned in the introduction, the fact that G can vary with time is not a new idea. P. Dirac in 1937, and later that year, P. Jordan, were both convinced that G has a cosmological origin, and more specifically, that G decreases with an increase in cosmological time. Dirac suggested that G is proportional to $t^{-1}$ where "t" is cosmological time while Jordan believed that $\dot{G}/G = -H$. Jordan also introduced a scalar field to model G, as we will likewise show.

If we accept the identification of $G/G_0$ with $a^{-\alpha}$, as was done in equation (2-4), then by equation (2-3), the matter and radiation mass densities must also scale by this factor. In fact, the following modifications have to be introduced for these terms.

$$\Omega_{RAD} \, a^{-4} \quad \rightarrow \quad \Omega_{RAD} \, a^{-4-\alpha} \tag{2-8a}$$

$$\Omega_{MATTER} \, a^{-3} \quad \rightarrow \quad \Omega_{MATTER} \, a^{-3-\alpha} \tag{2-8b}$$

However, we keep in mind that $\alpha$ currently, is only about .06 in value, very small compared to -3 or -4. Furthermore $\alpha$ will not vary much for most of the evolution of the universe; as we shall see; it is only in the very earliest phases in the universe where $\alpha$ changes its value appreciably. For $a \rightarrow 0$, it will turn out that $\alpha \rightarrow +1$, and $w \rightarrow -2/3$. The quintessence parameter, w, in our framework will never decrease below -2/3. For an opposing limit, $a \rightarrow \infty$, it will turn out that $\alpha \rightarrow 0$, and $w \rightarrow -1$, and we retrieve the concordance limit. Moreover, for most of the universe, $\alpha$ will have a relatively low value. When calculating look-back times and the age of the universe, however, as in section IV, this dependency of matter and radiation densities on $\alpha$ will be taken into account.

We next focus our attention on $\dot{G}/G$. We take the derivative of equation (2-4) resulting in

$$\dot{G}/G_0 = -a^{-\alpha} \, \dot{\alpha} \, \ln(a) - a^{-\alpha-1} \, \alpha \, \dot{a} \tag{2-9}$$

We have made use of the mathematical identity: $d/dx \, (f^{\,g}) = f^{\,g} *(dg/dx)* \ln(f) + f^{\,g-1} * g * df/dx$ recognizing that both "a" and "α" are functions of time in equation (2-4). We can divide equation (2-9) by equation (2-4), the left hand side by the left hand side, and the right hand side by the right hand side. This allows us to write

$$\dot{G}/G = -\dot{\alpha} \ln(a) - \alpha H \qquad (2\text{-}10)$$

In the present epoch, a =1, and the first term vanishes. We also have a good estimate for w in the current epoch, namely $w_0 = -.98$. Hence $\alpha \equiv 3(1+w) = .06$. This we substitute into equation (2-10) to obtain

$$(\dot{G}/G)\,|_0 = -.06 \, H_0 \qquad (2\text{-}11)$$

This result is noteworthy because it shows us that a) $(\dot{G}/G)$ is proportional to Hubble's parameter as first proposed by Jordan, and b) it is negative as suggested by Dirac and Jordan. However, its value is less than the full Hubble value as originally proposed by Jordan. Jordan claimed that $(\dot{G}/G)\,|_0 = -H_0$. As far as we know, our estimate for $(\dot{G}/G)\,|_0$, as specified by equation (2-11), does not conflict with the latest observational bounds on $(\dot{G}/G)$. Our estimate for $\dot{G}/G|_0$ works out to be about $-4 * 10^{-12}$ yr$^{-1}$; the exact value is dependent on the value of $H_0$ ultimately chosen. Observational constraints require $(\dot{G}/G)$ to be less than about $.1 \, H_0$ and equation (2-11) does not contradict this bound. Furthermore, many observational tests are at the very limit of the estimate given above, making the prediction in equation (2-11) especially interesting from a testing point of view.

Equation (2-10) is an interesting formulation for $\dot{G}/G$ and yet, it is of limited value. First the relation (2-10) is a=1 centric, and thus, is not a good candidate for a cosmological equation. A cosmological equation cannot single out a particular spatial or temporal point in the universe, and equation (2-10) does just that in positioning itself around the present epoch, a=1. A second problem with equation (2-10) is that α needs to be expressed in terms of "a" (or vice versa) in order to get a specific dependency for G in terms of "a" or α. The quintessence parameter "w", and thus α, is a function of the scale parameter "a". We may know current values for $w_0$, and $\alpha_0$, but we do not know past or future values. Hence, equation (2-10) cannot be integrated. In the next section, we will advance two separate parametrizations for G(a). This will allow us to specify a particular evolution for G in terms of scale parameter, "a".

Since equation (2-10) is of limited value, we turn instead to equation (2-3). We take the square root of both sides to obtain

$$H/H_0 = (G/G_0)^{1/2} * (\Omega_{RAD} \, a^{-4} + \Omega_{MATTER} \, a^{-3} + \Omega_\Lambda)^{1/2} \qquad (2\text{-}12)$$

We next differentiate with respect to time. This gives

$$\dot{H}/H_0 = 1/2 \, (G/G_0)^{-1/2} \, \dot{G}/G_0 \, (\Omega_{RAD} \, a^{-4} + \Omega_{MATTER} \, a^{-3} + \Omega_\Lambda)^{1/2} + 1/2 \, (G/G_0)^{1/2} \, (\Omega_{RAD} \, a^{-4} +$$
$$\Omega_{MATTER} \, a^{-3} + \Omega_\Lambda)^{-1/2} \, (-4 \, \Omega_{RAD} \, a^{-5} - 3 \, \Omega_{MATTER} \, a^{-4}) \, \dot{a} \qquad (2\text{-}13)$$

A dot over any physical quantity will always indicate a derivative is to be taken with respect to time. We divide the left hand side of equation (2-13) by the left hand side of (2-12); we do the same thing on the right hand side. After some simplification, we obtain the result:

$$\dot{H}/H = \tfrac{1}{2}\,\dot{G}/G - \tfrac{1}{2}\,H\,(4\Omega_{RAD}\,a^{-4} + 3\Omega_{MATTER}\,a^{-3})/(\Omega_{RAD}\,a^{-4} + \Omega_{MATTER}\,a^{-3} + \Omega_\Lambda) \quad (2\text{-}14)$$

This equation can be analyzed. In the limit where a→0, we have a radiation dominated universe where equation (2-14) reduces to

$$\dot{H}/H = 1/2\,\{\dot{G}/G - 4H\} \qquad \text{(radiation dominated)} \qquad (2\text{-}15a)$$

In the matter-dominated era, $\Omega_{MATTER}$ prevails, and equation (2-14) simplifies to

$$\dot{H}/H = 1/2\,\{\dot{G}/G - 3H\} \qquad \text{(matter dominated)} \qquad (2\text{-}15b)$$

In addition, in the dark energy dominated era, where only $\Omega_\Lambda$ survives, equation (2-14) gives

$$\dot{H}/H = 1/2\,\{\dot{G}/G\} \qquad \text{(dark energy dominated)} \qquad (2\text{-}15c)$$

In the present epoch, we can estimate a value for $\dot{H}/H$ using equation (2-14).

$$\dot{H}/H = \tfrac{1}{2}\,(\dot{G}/G)\,|_0 - \tfrac{1}{2}\,H_0\,(4\Omega_{RAD} + 3\Omega_{MATTER})$$

$$= \tfrac{1}{2}\,(-.06\,H_0) - \tfrac{1}{2}\,(.927\,H_0)$$

$$= -.4935\,H_0 \qquad (2\text{-}16)$$

In equation (2-16), we have made use of $\Sigma\,\Omega_i = 1$ in the first line. We have also substituted the values for $(\Omega_{RAD}, \Omega_{MATTER}) = (8.3*10^{-5}, .3089)$ to obtain the second line. The $\dot{G}/G$ terms in equations (2-15a,b,c) and equation (2-16) are new. We note that if $\dot{G}/G = 0$, then the result in equation (2-16) would change slightly, to $-.4635\,H_0$. Equation (2-16), therefore, is reasonably close even though we are assuming that $\dot{G}/G \neq 0$.

W next consider the acceleration parameter, $\ddot{a}$. Using the definition of H as $H \equiv \dot{a}/a$, it can be shown quite generally that

$$\dot{H}/H = \ddot{a}/\dot{a} - H \qquad (2\text{-}17)$$

We can revisit equations (2-15a,b,c) and equation (2-16) with this in mind. Substituting equation (2-17) into each of these equations gives the following results

$$\ddot{a}/\dot{a} = \tfrac{1}{2}\,\dot{G}/G - H \qquad \text{(radiation dominated)} \qquad (2\text{-}18a)$$

$$\ddot{a}/\dot{a} = \tfrac{1}{2}\,\dot{G}/G - \tfrac{1}{2}\,H \qquad \text{(matter dominated)} \qquad (2\text{-}18b)$$

$$\ddot{a}/\dot{a} = \tfrac{1}{2}\,\dot{G}/G + H \qquad \text{(dark energy dominated)} \qquad (2\text{-}18c)$$

Furthermore, in the present epoch,

$$\ddot{a}/\dot{a}\,|_0 = .5065\,H_0 \qquad \text{(present epoch)} \qquad (2\text{-}19)$$

Since $\dot{a} > 0$, we see very clearly that $\ddot{a}$ is positive in the present epoch. If $\dot{G}/G = 0$, then the result in equation (2-19) is modified slightly and increases to $.5365\,H_0$. Both values, however, are comparable and hence the $\dot{G}/G \neq 0$ does not alter the present rate of cosmic expansion appreciably.

A standard result in cosmology relates the cosmological constant, $\Lambda$, to the mass density associated with dark energy. By construction,

$$\Lambda = 8\pi G \, \rho_\Lambda/c^2 \qquad (2\text{-}20)$$

From this, it follows that

$$\dot{\Lambda}/\Lambda = \dot{G}/G + \dot{\rho}_\Lambda/\rho_\Lambda \qquad (2\text{-}21)$$

However, by equation (2-5), we can show that $\dot{\rho}_\Lambda/\rho_\Lambda = \dot{G}/G$. Thus, equation (2-21) reduces to

$$\dot{\Lambda}/\Lambda = 2\,\dot{G}/G \qquad (2\text{-}22)$$

Finally, starting from equations (2-22), and (2-21), it can be demonstrated that the following relations hold

$$\Lambda/\Lambda_0 = (G/G_0)^2 = (\rho_\Lambda/\rho_{\Lambda 0})^2 \qquad (2\text{-}23)$$

From equation (2-23) we see that if $G = G_0$, then $\Lambda = \Lambda_0$ and $\rho_\Lambda = \rho_{\Lambda 0}$. These equalities are assumed in the concordance model, but not in this paper.

A specific model for $G = G(a)$ has not been given. Two parametrizations will be given in the next section. Nevertheless, from equation (2-23), it is clear that should $G/G_0 \gg 1$, then $\Lambda/\Lambda_0 \ggg 1$. We have indicated how the cosmological constant fine-tuning problem is to be explained. In the distant past, both $G$ and $\rho_\Lambda$ were very, very large in relation to present values. This increased the value for the cosmological constant, $\Lambda$, significantly at very high temperatures. In section V, we will make plausible that $G$ was about twenty orders of magnitude greater than the current value. Therefore, by equation (2-23), we will have over a 40-fold increase in $\Lambda$, over present value. In section V, we will stop well short of the Planck scale, as we will give arguments for why gravity must have switched off at a scale of about $7 * 10^{21}$ K. This is appreciably less than the Planck Temperature of $1.42 * 10^{32}$ K, which assumes a constant value for $G$ throughout. Because $G/G_0$ will never increase beyond $4 * 10^{20}$, we will never approach a $10^{122}$ increase in cosmological constant using equation (2-23).

### III   *TWO SPECIFIC PARAMETRIZATIONS FOR G(a)*

We have seen that $\alpha$ in equation (2-10) cannot be determined unless we specify a function for $G(a)$. Moreover, if $\alpha$ cannot be ascertained, neither can the quintessence parameter, "w", because of our definition $\alpha \equiv 3(1+w)$. In this section, we give two specific models for $G(a)$. Both are one-dimensional parametrizations, depending in effect only on the scale parameter, "a". The scale parameter, "a" is a measure of temperature because of the relationship, $a = (1+z)^{-1} = T_0/T$ where $T_0$ equals 2.725 K, and T is the CMB temperature at any other redshift z. We feel it is more meaningful to parametrize $G$ according to background temperature (energy), versus, for example, cosmological time. Cosmic conditions in the universe depend specifically on the background temperature and not on time per se. Both parametrizations which we are about to introduce have great flexibility in accommodating a wide range of $G$ values, and both are relatively simple. Whether they have any physical relevance remains to be seen. However, we can draw some general conclusions using these very basic models. Remarkably both lead to essentially similar results, both qualitatively and quantitatively, even though they are very different formulations for G(a). We have reasons for considering the above models, which go beyond the scope of this paper. Until then we consider these models to be "toy models".

The first parametrization is motivated by a charging capacitor; we can think of $G^{-1}(a)$ as a gravitational charge, which builds up over cosmological time. The lower the background temperature, the larger $G^{-1}$ becomes allowing for weaker gravitational coupling between masses. We call this model A and the underlying equation reads

$$G^{-1} = G_\infty^{-1} (1 - e^{-x}) \qquad (3-1)$$

In equation (3-1), "x" is defined as $x \equiv b/T = a\, b/T_0$ where "b" is a constant to be determined having units of degrees Kelvin, and "a" is our scale parameter. In the present epoch, "a" = 1, and thus, $x_0 \equiv b/T_0$. In equation (3-1), $G_\infty^{-1}$ is the saturation value of $G^{-1}$, applicable in the limit where the CMB temperature approaches zero, or equivalently, when "a" approaches infinity.

The second parametrization is motivated by magnetism. We treat $G^{-1}(a)$ as an order parameter, which vanishes at high energies (temperatures). At lower temperatures long-range correlations emerge which causes an alignment of sorts to produce inverse gravity. This we call model B and in this model,

$$G^{-1} = G_\infty^{-1} L(x)$$
$$= G_\infty^{-1} [\coth(x) - 1/x] \qquad (3-2)$$

In equation (3-2), L(x) is the Langevin function, defined by the equation $L(x) \equiv [\coth(x) - 1/x]$. As before $x \equiv b/T = a\, b/T_0$ where "b" is a constant to be determined, having units of degrees Kelvin, and $x_0 \equiv b/T_0$. In equation (3-2), $G_\infty^{-1}$ is a different saturation value for $G^{-1}$, but defined in the same way. In the limit where T approaches zero, $G^{-1}$ approaches $G_\infty^{-1}$. Plotting $G^{-1}(a)$ as a function of "x" gives similar behavior in both models. Graph 5a in appendix C is a preview of the functions plotted as a function of scale parameter "a" up to "a" = 1. Both are well behaved at both high and low temperatures, as we shall see. At very high temperatures in particular, it will be shown that $G^{-1}$ is directly proportional to T in both models, but only in this limiting case.

Using equation (3-1), we can show that

$$(G/G_0)|_A = (1 - e^{-x_0})/(1 - e^{-x}) \qquad (3-3)$$

For model B, equation (3-2) applies and we have correspondingly,

$$(G/G_0)|_B = (\coth(x_0) - 1/x_0)/(\coth(x) - 1/x) = L(x_0)/L(x) \qquad (3-4)$$

In both equations, it is to be understood that the temperature T marks a particular cosmological epoch. At the onset of $G^{-1}$, we will also have a very specific temperature, which we will call $T_C$, the Curie temperature.

To make progress with these parametrizations, the constant "b" needs to be determined. We know that at present, $(\dot{G}/G)|_0 = -.06\, H_0$, as is indicated by equation (2-11). We will use this equality to fix the "b" value for both models A and B. We start with model A. Take the derivative of equation (3-3) which yields

$$(\dot{G}/G_0)|_A = -[(1 - e^{-x_0})/(1 - e^{-x})^2](e^{-x})\dot{x} \qquad (3\text{-}5)$$

Divide this by equation (3-3), left hand side by left hand side, and right hand side by right hand side, to give

$$(\dot{G}/G)|_A = -[(e^{-x})/(1 - e^{-x})]\dot{x}$$

$$= -xH/(e^x - 1) \qquad (3\text{-}6)$$

Remember that $\dot{x} = \dot{a} x_0 = (\dot{a}/a) x = H x$. Next, specialize to the current epoch. In this limit, equation (3-6) becomes

$$(\dot{G}/G_0)|_{A0} = -x_0 H_0/(e^{x_0} - 1) \qquad (3\text{-}7)$$

This can be compared to equation (2-11), from which it follows that equation (3-7) can be written as

$$-.06 H_0 = -x_0 H_0/(e^{x_0} - 1) \qquad (3\text{-}8)$$

The $H_0$ cancels and a numerical solution can be found to fix the parameter $x_0$. For equation (3-8) to be satisfied, we must uniquely choose $x_0 = 4.28$. Hence $b = x_0 T_0 = 4.28 * 2.725 = 11.663$ K. Summarizing, for model A, we therefore require that

$$x_0 = 4.28, \qquad b = 11.663 \text{ K} \qquad \text{(model A)} \qquad (3\text{-}9)$$

This fixes our parametrization for model A.

For model B, we proceed similarly. Take the derivative of equation (3-4) to obtain

$$(\dot{G}/G_0)|_B = +[(\coth(x_0) - 1/x_0)/(\coth(x) - 1/x)^2](x\,\text{csch}^2 x - 1/x)\dot{x}/x \qquad (3\text{-}10)$$

We divide the left hand side of equation (3-10) by the left hand side of equation (3-4); do the same on the right hand side. In this way we obtain

$$(\dot{G}/G)|_B = [(x\,\text{csch}^2 x - 1/x)/(\coth(x) - 1/x)] H \qquad (3\text{-}11)$$

As before, we utilized the identity, $\dot{x} = H x$. We next specialize (3-11) to the present epoch, which gives

$$(\dot{G}/G)|_{B0} = [(x_0\,\text{csch}^2 x_0 - 1/x_0)/(\coth(x_0) - 1/x_0)] H_0 \qquad (3\text{-}12)$$

We compare this to equation (2-11). If equation (2-11) is substituted into (3-12), it turns out that

$$-.06 H_0 = [(x_0\,\text{csch}^2 x_0 - 1/x_0)/(\coth(x_0) - 1/x_0)] H_0 \qquad (3\text{-}13)$$

Again, $H_0$ cancels and a numerical solution can be found. We find that left and right hand sides of equation (3-13) are equal if and only if we choose $x_0 = 17.67$. Thus the "b" in model B has also been uniquely determined because we know that $b = x_0 T_0 = 17.67 * 2.725 = 48.15$ K. Summarizing, for model B

$$x_0 = 17.67, \quad b = 48.15 \text{ K} \quad \text{(model B)} \quad (3\text{-}14)$$

This determines our parametrization for model B. We note that the temperatures indicated in equations (3-9) and (3-14) are not particularly high.

We have now specified both functions for $G^{-1}(a)$. For model A, we use equation (3-3) with the parameters given in (3-9) substituted. Explicitly,

$$(G/G_0)|_A = (1 - e^{-4.28})/(1 - e^{-4.28 a}) \quad \text{(model A)} \quad (3\text{-}15)$$

For model B, we do the same. We use equation (3-4) with the parameters given in equation (3-14) substituted. This allows us to write

$$(G/G_0)|_B = [(\coth(17.67) - 1/17.67)/(\coth(17.67 a) - 1/(17.67 a))] \quad \text{(model B)} \quad (3\text{-}16)$$

For any epoch, G can now be calculated using either equation (3-15) or (3-16). These functions depend only on the cosmic scale parameter, "a", or equivalently the CMB temperature.

We next determine $\dot{G}/G$ for both models. For model A, we use equation (3-6) with $x = a\, x_0 = (a\, 4.28)$ substituted. For model B, we use equation (3-11) with $x = a\, x_0 = (a\, 17.67)$ substituted. Given any scale factor, we can thus determine a specific value for "x", and thus find $(\dot{G}/G)|_A$ and $(\dot{G}/G)|_B$ as a function of "a". We thus have $G/G_0$ and $(\dot{G}/G)$ for both models. Equations (3-15) and (3-16) give $G/G_0$. In addition, equations (3-6) and (3-11), with the appropriate ($a\, x_0$) values substituted, give $(\dot{G}/G)$.

Finally, we also wish to calculate the quintessence parameter, w, for models A and B, as well as determine α for models A and B. We know that equation (2-4) holds. Therefore, it follows that

$$\ln(G/G_0) = -\alpha \ln(a) \quad (3\text{-}17)$$

However, we have specific values for $G/G_0$ as a function of "a". Hence,

$$\alpha = -\ln(a)/\ln(G/G_0) \quad (G \neq G_0) \quad (3\text{-}18)$$

Alternatively,

$$3(1+w) = -\ln(a)/\ln(G/G_0) \quad (G \neq G_0) \quad (3\text{-}19)$$

Equation (3-18) allows us to determine α as a function of the scale parameter "a", whereas equation (3-19) allows us to calculate "w".

We can now summarize the results. These are presented in table form, table I in appendix A for various values of "a". In this table we calculate $G/G_0$, $(\dot{G}/G)$, α, and w for both models, A and B. The scale parameter is indicated under column 1. The corresponding $(G/G_0)|_A$, $(G/G_0)|_B$, $(\dot{G}/G)|_A$, $(\dot{G}/G)|_B$ values are specified under columns 2, 3, 4 and 5, respectively. The "α" and "w" values are specified under columns 6, 7, 8 and 9. Columns 6 and 7 give the "α" values for models A and B, respectively, whereas columns 8 and 9 are reserved for the "w" values for models A and B, respectively. The table is broken up in four parts. In part one, we cover the range where "a" equals 1 through to .1. In part two, "a" values in the range .1 to .01 are considered. In part three, "a" is allowed to run through the values from .01 to .001. In addition, in part four, we consider future values for "a". In this part, "a" will start at 1 and move up to 10, in increments of one unit per row. So, in the first three parts, we are going progressively back in time, while in the 4th part, we are moving forward in cosmic evolution.

We also present graphs for the quantities calculated above. These visuals are given in appendix B, and the values correspond to the entries specified in table I. We present the graphs in a certain order. First model A is always compared with model B, quantity with corresponding quantity. This is done such that we can visually compare the difference between models. On the horizontal axis, we always plot the scale parameter "a". On the vertical axis, we plot $G/G_0$, ($\dot{G}/G$), "α" and "w" for both models A and B. Graphs 1a, 1b, 1c and 1d give these values for "a" in the range from 1 to .1. Graphs 2a, 2b, 2c, and 2d do the same for "a" values in the range from .1 and .01. Graphs 3a, 3b, 3c, and 3d are reserved for "a" values in the range from .01 to .001. And finally, graphs 4a, 4b, 4c, and 4d give the values outlined above for "a" in the range from 1 to 10. Therefore, graphs 1 refer to part 1 in the table, graphs 2 refer to part 2 in the table, graphs 3 to part 3 in the table, and graphs 4 refer to part 4 in the table.

Upon comparing the numerical values between models A and B, we notice a remarkable similarity. Model B is somewhat more conservative in that $G/G_0$ does not increase quite as dramatically as in model A when we go back in time towards higher temperatures. Nevertheless, the order of magnitude estimates seem to be on a par right up to and including the very early universe. Both models lead to essentially the same results, even though the two parametrizations are distinct from each other. We remark that these functionalities for $G^{-1}$ in terms of "a" depends critically on our original choice for w, namely that specified by equation (1-1). This determined equation (2-11), which furthermore allowed us to fix the parameter "b" in both models. If we had chosen a value closer to -1 for w, then there would be little to no difference between the ΛCDM concordance model, and our models A and B. *Our parametrizations deviate from the ΛCDM model precisely because we do not set w equal to -1, a-priori*. A higher or lower value for "w" at present will dramatically affect the evolution of $G/G_0$.

We also note that in the limit of low T, the "w" values automatically approach -1, and α approaches zero. *Thus, the ΛCDM model is approached in both our models in the low temperature limit*. At very high temperatures, on the other hand, the quintessence parameter, w, approaches a value of -2/3 in both models. This will give a value for α equal to unity.

Both models A and B have the correct limits for an order parameter, $G^{-1}$ (a). This we will now show. First, quite generally, irrespective of the model employed, it is to be noticed that the following general identity holds:

$$d/dt\ (G^{-1})/\ G^{-1} = -\ G^{-2}\ \dot{G}/G^{-1} = -\ \dot{G}/G \qquad (3\text{-}20)$$

Upon using equation (3-20), it is straightforward to show that

$$d/dt\ (G^{-1})/\ (G^{-1}) = -\ \dot{G}/G$$

$$= x\ H/\ (e^x - 1) \qquad \text{(model A)} \qquad (3\text{-}21)$$

And, for model B,

$$d/dt\ (G^{-1})/\ (G^{-1}) = -\ \dot{G}/G$$

$$= - [(x \operatorname{csch}^2 x - 1/x)/ (\coth x - 1/x)] H \quad \text{(model B)} \quad (3\text{-}22)$$

In equation (3-21), we have made use of equations (3-6). In addition, for equation (3-22), equation (3-11) was employed.

We can consider the limit where a→0. In this limit, $x \equiv (a x_0) \to 0$. Therefore, for small values of x, $e^x \cong (1+x)$, and equation (3-21) reduces to

$$\text{Lim } a \to 0 \quad d/dt \ (G^{-1})/ (G^{-1}) = H \quad \text{(model A)} \quad (3\text{-}23)$$

A similar result holds for equation (3-22). For small values of x, a power series expansion yields

$$(x \operatorname{csch}^2 x - 1/x) = -x/3 + x^3/15 - 2x^5/189 + \ldots \quad (3\text{-}24a)$$

$$(\coth x - 1/x) = x/3 - x^3/45 + 2x^5/945 + \ldots \quad (3\text{-}24b)$$

Keeping only terms to first order in x, equation (3-22) reduces to

$$\text{Lim } a \to 0 \quad d/dt \ (G^{-1})/ (G^{-1}) = H \quad \text{(model B)} \quad (3\text{-}25)$$

In this limit of very small "a", it is clear that $d (G^{-1})/ (G^{-1}) = da/ a = - dT/ T$, for both models A and B. At very high temperatures, i.e., very low "a" values, it follows that $(G_1^{-1}/ G_2^{-1}) = G_2/ G_1 = a_1/ a_2 = T_2/ T_1$.

The other extreme is the limit where a→∞, which means going forward in time starting from the present epoch. In this interesting case, the reader will notice that the entries in our table I indicate a saturation value for both models A and B. In fact, as can be read off the table, specifically the entries under columns 2 and 3, we find in the limit of large "a"

$$(G_\infty) |_A = .986 \ G_0 \quad \text{(model A; large "a")} \quad (3\text{-}26)$$

$$(G_\infty) |_B = .949 \ G_0 \quad \text{(model B; large "a")} \quad (3\text{-}27)$$

Hence,

$$(G_\infty^{-1}) |_A = 1.014 \ G_0^{-1} \quad \text{(model A; large "a")} \quad (3\text{-}28)$$

$$(G_\infty^{-1}) |_B = 1.054 \ G_0^{-1} \quad \text{(model B; large "a")} \quad (3\text{-}29)$$

These are the saturation values to be used in equations (3-1) and (3-2), respectively. At present, $G^{-1}$ is varying very slowly because we are already close to saturation. The value $G^{-1}$ will eventually stop increasing in both models. For model A, this occurs already at roughly a ≈ 2. For model B, $G^{-1}$ becomes virtually a constant at a ≈ 10. See table I, and the graphs given in appendix B, specifically graphs 4a and 4b. We are late in the evolution of $G^{-1}$ in the present epoch, and Newton's constant, $G_0$, will not decrease much further as equations (3-26) and (3-27) are the limits calculated by our models.

Before we leave this section, we give another table; table II, in appendix C. Here we calculate $G^{-1}/G_0^{-1}$, and $\{d (G^{-1})/dt\}/ G^{-1}$ as a function of "a", but moving forward in time starting in the distant past. We start with "a" = .05 and work our way up to the present day where "a"=1. This is more natural as we are proceeding from higher energies to lower ones, from a time in the distant past to the present. Keep in mind that $G^{-1}$ is our order parameter, and not G. For $G^{-1}$, we use equations (3-1) and (3-2) with the parameters (3-9) and (3-14) substituted, respectively. Equations (3-1) with (3-9) hold for model A while equations (3-2) with (3-14) hold for model B. Furthermore, we can make use of equation (3-20) for $\{d(G^{-1})/dt\}/G^{-1}$. This is given in units of H, Hubble's parameter. This is just another way of representing what was said thus far. However, in this formulation, we clearly see the development of the order parameters and the differences between the specific evolutions for models A and B.

We highlight this difference between models A and B more explicitly using graphs. In appendix C, we present two graphs, graph 5a and 5b. Graph 5a gives $G^{-1}/G_0^{-1}$ for our models A and B. Graph 5b, on the other hand, gives $\{d(G^{-1})/dt\}/G^{-1}$ for both models, in units of H. In both models, as $G^{-1}/G_0^{-1}$ increases, $\{d(G^{-1})/dt\}/G^{-1}$ decreases. At present, $\{d(G^{-1})/dt\}/G^{-1}$ hardly changes at all. Again, this is just another way of presenting what was determined previously.

## IV    *COSMIC TIME EVOLUTION FOR $G^{-1}(a)$*

In the conventional picture, the $\Lambda$CDM model, where $\dot{G}/G = 0$, and w = -1, we know that a(t) is proportional to $t^{1/2}$ for a radiation dominated universe, a(t) is proportional to $t^{3/2}$ for a matter dominated universe, and a(t) is proportional to $e^{Ht}$ for a dark energy dominated universe. If $\dot{G}/G \neq 0$, however, the time dependency is more complicated. This we now consider.

To be specific, we resort to our two parametrizations, model A and model B given by equations (3-1 and 3-2), respectively. First we review the steps in the $\Lambda$CDM model where $\dot{G}/G = 0$, and w = -1. Utilizing equation (2-2), and recognizing that H = $\dot{a}/a$, we can see that

$$\dot{a}/a = H_0 (\Omega_{RAD}\, a^{-4} + \Omega_{MATTER}\, a^{-3} + \Omega_\Lambda)^{1/2} \tag{4-1}$$

Thus

$$dt = H_0^{-1} (\Omega_{RAD}\, a^{-4} + \Omega_{MATTER}\, a^{-3} + \Omega_\Lambda)^{-1/2} \; da/a \tag{4-2}$$

Integrating gives

$$(t_0 - t) = H_0^{-1} \int_a^1 (\Omega_{RAD}\, a^{-4} + \Omega_{MATTER}\, a^{-3} + \Omega_\Lambda)^{-1/2}\, da/a \tag{4-3}$$

, where $(t_0 - t)$ is the look-back time. Setting t = 0 corresponds to going back to the beginning of cosmological time, where a = 0. If we go forward in time, we can define a look-forward time where the limits of integration are reversed, and the left hand side is replaced by $(t - t_0)$. Both integrals are performed using numerical integration once the $\Omega_{RAD}$, $\Omega_{MATTER,}$ and $\Omega_\Lambda$ values have been substituted. In the above equation, we can take $H_0$ equal to 67.74 km/(s Mpc) = $2.195*10^{-18}$ $s^{-1}$ = $(14.44 * 10^9)^{-1}$ $yr^{-1}$, and the density parameters can be chosen as ($\Omega_{RAD}$, $\Omega_{MATTER}$, $\Omega_\Lambda$) = ($8.3*10^{-5}$, .3089, .6911). This would conform to the parameters suggested by the Planck VIII collaboration. Equation (4-3) will give us precisely the age of the universe, $t_0$ = 13.8 Gyr, if we set t = 0 on the left hand side and "a"= 0 on the right.

We note that in the very early universe, radiation dominates due to the high value of $a^{-4}$. In this instance, the second and third terms on the right hand side of equation (4-3) are negligible, and we obtain the customary $t^{1/2}$ dependency for a(t) at high temperatures. For a matter-dominated universe, the $\Omega_{MATTER}$ term dominates in equation (4-3), because the first and third terms are small for intermediate "a" values. Here it is easy to show that a(t) is proportional to $t^{3/2}$. Finally, for a dark energy dominated universe, the third term takes over within the integral of equation (4-3). In this instance, we have a $e^{Ht}$ dependency for a(t). In table II, column 2, in appendix C we have calculated specific look-back times as well as look-forward times for the $\Lambda$CDM model as a function of scale parameter, "a", which is indicated in column 1. We made use of equations (4-3) with the density parameter coefficients inserted. Numerical integration was performed using an on-line integrator, integral-calculator.com.

For models A and B, both equations (4-2) and (4-3) are modified. For model A, the counterpart to equation (4-3) is

$$(t_0 - t) = H_0^{-1} \int_x^{x_0} [(1-e^{-x})/(1-e^{-x_0})]^{1/2} (\Omega'_{RAD} x^{-2} + \Omega'_{MATTER} x^{-1} + \Omega_\Lambda x^2)^{-1/2} dx \quad (4-4)$$

To show this we start with equation (3-3). We substitute equation (3-3) into equation (2-12) to obtain

$$\dot{a}/a = H_0 [(1-e^{-x_0})/(1-e^{-x})]^{1/2} (\Omega_{RAD} a^{-4} + \Omega_{MATTER} a^{-3} + \Omega_\Lambda)^{1/2} \quad (4-5)$$

From this, it follows that

$$dt = H_0^{-1} [(1-e^{-x})/(1-e^{-x_0})]^{1/2} (\Omega_{RAD} a^{-4} + \Omega_{MATTER} a^{-3} + \Omega_\Lambda)^{-1/2} da/a \quad (4-6)$$

However, $x = a\, x_0$. Therefore $dx = da\, x_0$ and $da/a = dx/x$. Furthermore $a^{-4} = (x/x_0)^{-4}$, and $a^{-3} = (x/x_0)^{-3}$. We redefine $\Omega_{RAD}, \Omega_{MATTER}$ as $\Omega'_{RAD}, \Omega'_{MATTER}$, where

$$\Omega'_{RAD} \equiv \Omega_{RAD}\, x_0^4 = (8.3 * 10^{-5})(4.28)^4 = .02785 \quad (4-7a)$$

$$\Omega'_{MATTER} \equiv \Omega_{MATTER}\, x_0^3 = (.3089)(4.28)^3 = 24.2186 \quad (4-7b)$$

This allows us to rewrite equation (4-6) as

$$dt = H_0^{-1} [(1-e^{-x})/(1-e^{-x_0})]^{1/2} (\Omega'_{RAD} x^{-4} + \Omega'_{MATTER} x^{-3} + \Omega_\Lambda)^{-1/2} dx/x$$

$$= H_0^{-1} [(1-e^{-x})/(1-e^{-x_0})]^{1/2} (\Omega'_{RAD} x^{-2} + \Omega'_{MATTER} x^{-1} + \Omega_\Lambda x^2)^{-1/2} dx \quad (4-8)$$

Equation (4-4) follows from relation (4-8). Since $x_0 = 4.28$, we can evaluate the constant, $(1-e^{-x_0})^{1/2} = .9931$.

For model B, the analogue of equation (4-3) is

$$(t_0 - t) = H_0^{-1} \int_x^{x_0} [(\coth(x)-1/x)/(\coth(x_0)-1/x_0)]^{1/2} (\Omega'_{RAD} x^{-2} + \Omega'_{MATTER} x^{-1} + \Omega_\Lambda x^2)^{-1/2} dx \quad (4-9)$$

We follow the same steps as before. We start with equation (3-4) and substitute this into equation (2-12). This gives

$$\dot{a}/a = H_0 \, [(\coth(x_0)-1/x_0)/ (\coth(x)-1/x)]^{1/2} \, (\Omega_{RAD} \, a^{-4} + \Omega_{MATTER} \, a^{-3} + \Omega_\Lambda)^{1/2} \quad (4\text{-}10)$$

From this, it follows that

$$dt = H_0^{-1} \, [(\coth(x)-1/x)/ (\coth(x_0)-1/x_0)]^{1/2} \, (\Omega_{RAD} \, a^{-4} + \Omega_{MATTER} \, a^{-3} + \Omega_\Lambda)^{-1/2} \, da/a \quad (4\text{-}11)$$

However, $da/a = dx/x$. Also, $a^{-4} = (x/x_0)^{-4}$ and $a^{-3} = (x/x_0)^{-3}$, as before. If we redefine our density parameters as follows:

$$\Omega'_{RAD} \equiv \Omega_{RAD} \, x_0^4 = (8.3 * 10^{-5}) (17.67)^4 = 8.0914 \quad (4\text{-}12a)$$

$$\Omega'_{MATTER} \equiv \Omega_{MATTER} \, x_0^3 = (.3089)(17.67)^3 = 1.7042 * 10^3 \quad (4\text{-}12b)$$

, then we can rewrite equation (4-11) as

$$dt = H_0^{-1} \, [(\coth(x)-1/x)/ (\coth(x_0)-1/x_0)]^{1/2} \, (\Omega'_{RAD} \, x^{-2} + \Omega'_{MATTER} \, x^{-1} + \Omega_\Lambda \, x^2)^{-1/2} \, dx \quad (4\text{-}13)$$

From this equation, it is clear how equation (4-9) follows. Because $x_0 = 17.67$, we can work out the constant in equation (4-13), namely, $(\coth(x_0)-1/x_0)^{1/2} = .9713$. As always, to obtain the look-forward times, we reverse the limits of integration in both equations (4-4) and (4-9), and substitute for the left hand side, $(t - t_0)$.

Now that we have equation (4-4) for model A, and correspondingly, equation (4-9) for model B, we can numerically evaluate the integrals. All coefficients are known except for x. The limits of integration for model A, are from $x = (a \, x_0) = (4.28 \, a)$ to $x_0 = 4.28$. This is to be used with equation (4-4). For model B, we use equation (4-9) where the limits are from $x = (a \, x_0) = (17.67 \, a)$ to $x_0 = 17.67$. The results of the numerical integrations are specified in table III in appendix D, under columns 3 and 4. These are calculated as a function of scale parameter "a", which is given under column 1. Column 3 holds for model A, and column 4 is valid for model B. These values can be compared to those values for the $\Lambda$CDM model, which we have listed under column 2. Upon comparison of the numerical results, most values of "a" give similar results. It is only when one gets to relatively low "a" values where one notices a real deviation.

A graph comparing the three models is illustrated in Graph 6a. One obvious difference between the models is the predicted age of the universe. The $\Lambda$CDM model gives a predicted age of $.9559 \, H_0^{-1} = .9559 \, (14.44 \text{Gyr}) = 13.8$ Gyr. Model A, by contrast, predicts a naïve value equal to $t_0 = .8688 \, H_0^{-1} = .8688 \, (14.44 \text{Gyr}) = 12.5$ Gyr. Moreover, model B predicts a third value equal to $t_0 = .9018 \, H_0^{-1} = .9018 \, (14.44 \text{Gyr}) = 13.0$ Gyr. Quite generally, the age of the universe is given by the expression, $t_0 = F(\Omega_{RAD}, \Omega_{MATTER}, \Omega_\Lambda) * H_0^{-1}$, where F is the so-called "age correction factor" [47]. The function F is specifically what we are plotting in Graph 6a. Its value depends on the density parameters chosen. Since we have made a specific choice, the values for F are determined as specified above. The age correction factors are (.9559, .8688, .9018) for the ($\Lambda$CDM, Model A, Model B) models, respectively, if we accept $\Omega_{RAD} = .000083$, $\Omega_{MATTER} = .3089$, and $\Omega_\Lambda = .6911$ as our input values. Seeing that the predicted ages of the universe in models A and B seem low, we have several options:

a) $H_0^{-1}$ has to be adjusted
b) The values for $\Omega_{RAD}$, $\Omega_{MATTER}$, and $\Omega_\Lambda$ have to be changed
c) The age is as specified, i.e., the universe is less than 13.8 Gyr old

    d) Some combination of the above.     (4-14)

We will discount option c) as the 13.8 Gyr age seems to be a well-established fact. The age of the oldest globular clusters certainly indicate an age in excess of 12.5 Gyr. The analysis of the acoustic peaks in WMAP and Planck satellite data also seem to preclude a lessor age for the universe. Therefore, we will discount option c). We will also ignore option d) as we are focused on most likely principle sources. This leaves options a) or b).

Option b) would require adopting a new set of values for $\Omega_{MATTER}$ and $\Omega_\Lambda$, other than the .3089 and .6911 values chosen, respectively. The $\Omega_{RAD}$ = .000083 value is not that critical for the numerical age determination of the universe. Adjusting $\Omega_{MATTER}$ and $\Omega_\Lambda$ values is a distinct possibility; it means changing the values towards a higher dark energy component, and a lessor dark matter contribution. Hence, we would be apt to dismiss option b) as well. However, we remark that if we had made a different choice for model A, namely, $\Omega_{MATTER}$ = .212, and $\Omega_\Lambda$ = .788, then the correction factor would match that of the $\Lambda$CDM result, namely, .9559. For model B, if we adjust $\Omega_{MATTER}$ to equal .249 with $\Omega_\Lambda$ equaling .751, then we would also match the $\Lambda$CDM correction factor exactly without any need for further fine-tuning. If we do not wish to change the values of the density parameters as determined by the Planck VIII collaboration, this leaves option a) as our best option.

Working within the framework of option a), we adjust our $H_0^{-1}$ values accordingly such that we reproduce 13.8 Gyr as the age of the universe. For model A, we demand specifically that

$$.9559\ H_0^{-1} = .8688\ H_{0A}^{-1} \qquad (4\text{-}15)$$

, where $H_{0A}$ is the value of the Hubble constant to be used for model A, and $H_0$ = 67.74 km/(s Mpc), the specified value as determined by the 2015 Planck XIII cosmological parameter collaboration. Solving equation (4-15) gives as a value for $H_{0A}^{-1}$

$$H_{0A}^{-1} = 15.99\ \text{Gyr} \qquad \text{or,} \qquad H_{0A} = 61.7\ \text{km/(s Mpc)} \qquad (4\text{-}16)$$

For model B, likewise, we can demand that

$$.9559\ H_0^{-1} = .9018\ H_{0B}^{-1} \qquad (4\text{-}17)$$

, where $H_{0B}$ is the value of Hubble's constant to be employed for model B, and $H_0$ is the established value as determined by the 2015 Planck XIII collaboration. Solving equation (4-17) gives a different value for $H_{0B}^{-1}$. We obtain in this instance,

$$H_{0B}^{-1} = 15.3\ \text{Gyr} \qquad \text{or,} \qquad H_{0B} = 63.9\ \text{km/(s Mpc)} \qquad (4\text{-}18)$$

The values indicated in equations (4-16) and (4-18) may seem low. However, we believe that they are entirely within of the range of observations.

We believe that these assignments for $H_{0A}$ and for $H_{0B}$ can be justified because $H_0$ is not that well determined, observationally. In fact, as of 2015, there is still a range of values, which can be assigned to $H_0$. As noted in reference[48], the Planck 2015 XIII cosmological parameter collaboration, there is a discrepancy (referred to as "tension") between the 9 year WMAP determination of $H_0$ and the newer Planck satellite limit on $H_0$. The latest Planck determination for $H_0$ is less than that determined previously by WMAP. Furthermore, analyzing other data, such as Cepheid Variables, give limits for $H_0$ as high as [49] 74.3 +/- 2.6 km/(s Mpc) [Friedmann, et.al. 2012] and as low as [50] 63.7 +/- 2.3 km/(s Mpc) [Tammann & Reindl, 2013]. These are not to be ruled out. The latter limit is interesting because both our models A and B fall clearly within this range. Model B is almost a perfect match, whereas model A is within the lower limit. Hence, if we normalize $H_0$ appropriately as was done in equations (4-14) and (4-16), then we believe we can keep the 13.8 Gyr age of the universe intact, and still stay within our model parametrizations. No revision in either $\Omega_{MATTER}$ or $\Omega_\Lambda$ is needed. The normalized results for models A and B are presented in table IV of appendix D. The corresponding graph is given in Graph 6b. Upon comparison of the entries in columns 3 and 4 with those of column 2, the $\Lambda$CDM model, we see agreement. Models A and B indicate slightly larger look-back times in the latter epochs (large "a"). Nevertheless, eventually, the $\Lambda$CDM model catches up with the same look-back times at a lower "a". We speculate that the larger look-back times in the latter epochs give the unanticipated acceleration of the universe. Ultimately, it is connected to a weakening G in the latter epochs.

We close this section by noting that if we wish to compare look-back times between models A and B, and $\Lambda$CDM, it may turn out that "w" has a different value at present than the one selected, $w_0 = -.98$. This is almost certainly true if space is not flat. If space is curved, then instead of using equation (1-1) for w, we should be using equation (1-2), or some variation thereof, as our approximation for "w". Moreover, if we accept the value for "w" listed in (1-2), then it would be very difficult to differentiate our models A and B from the $\Lambda$CDM model. In fact, our models A and B would be almost identical to the $\Lambda$CDM model in terms of predictions because "w" is so close to -1. For other "w" values in a flat space, the look-back times would also have to be re-worked. We bring this up only to show that there is greater flexibility than is indicated by our conditions (4-14), in either dismissing or accepting a variable G assumption given a specified age.

## V  ESTIMATING THE ONSET OF GRAVITY, $G^{-1}(a_C)$

Our parametrizations, given by equations (3-1) and (3-2), assume that $G^{-1}(a)$ is an order parameter, which approaches a constant value at low energies, i.e., at low CMB temperatures. The values measured today for $G_0$ are fairly close to these constant saturation values, $G_\infty$ as is indicated by equations (3-28) and (3-29). This can also be seen in graph 5a in appendix C where a saturation value is clearly visible. At some time in the distant past, however, at low enough "a", i.e., at sufficiently high CMB temperatures, $G^{-1}$ must have come into existence. In other words, gravity, as we know it, must have emerged. A small value for $G^{-1}$ must mean a large value for G. This ansatz was specifically introduced to solve the cosmic vacuum fine-tuning problem. See the discussion around equation (2-23). The big question now arises: can we estimate the onset of $G^{-1}$? Can we give a specific temperature, or equivalently, a specific cosmic scale, for G formation? We will call the critical temperature for G formation, $T_C = T(a_C)$, where $a_C$ is the scale parameter at inception of $G^{-1}$. $G(a_C)$ is thus the same as $G(T_C)$ because of the relation between scale parameter and CMB temperature, $a = (1+z)^{-1} = T_0/T$.

To answer this question, we first consider the Planck scale. At this scale, we supposedly have super grand unification where gravity, thermodynamics, and quantum mechanical fluctuations create a so-called "quantum foam" where virtual particles constantly interact and exchange energy with the underlying space-time continuum producing a turbulent vacuum froth. Neither space-time nor matter/energy will have a clear identity under those conditions. This is also the scale where our knowledge of physics breaks down because we cannot say anything intelligent beyond it, energy wise. The Planck scale invariably involves Newton's constant, G. In fact, the Planck length, the Planck time, the Planck mass, the Planck energy, etc., are all defined explicitly in terms of G. To be specific, the Planck length $L_P \equiv (\hbar G/c^3)^{1/2} = 1.62 * 10^{-35}$ m. The Planck mass is given by $M_P \equiv (\hbar c/G)^{1/2} = 2.18 * 10^{-8}$ kg. The Planck time is defined as $t_P \equiv (\hbar G/c^5)^{1/2} = 5.39 * 10^{-44}$ s. We have the Planck energy, $E_P \equiv (\hbar c^5/G)^{1/2} = 1.96 * 10^9$ J $= 1.225 * 10^{19}$ GeV. Then there is the Planck temperature, $T_P \equiv (\hbar c^5/G k_B^2)^{1/2} = 1.42 * 10^{32}$ K, etc. We see that all incorporate G, Newton's constant. The notable exception is the Planck charge $q_P \equiv (4\pi\varepsilon_0 \hbar c)^{1/2} = 1.88 * 10^{-18}$ C, which, interestingly, does not involve G explicitly. However, if G is not fundamental, then it can be argued that neither is the Planck scale. In fact, if we believe gravity to be a low energy phenomenological limit, then there is nothing significant about the Planck scale.

Nevertheless, there is one relationship in the above, which can prove useful. That is the Planck temperature, $T_P \equiv (\hbar c^5/G k_B^2)^{1/2}$. Instead of $T_P$, we replace the left hand side by $T_C$, the Curie temperature, which will signify the onset of gravity. Moreover, on the right hand side we recognize that G is also temperature dependent. This means that at the formation of G, we must have

$$T_C \equiv (\hbar c^5/G_C k_B^2)^{1/2} \qquad (5\text{-}1)$$

, where $G_C = G(T_C) = G(T_0/a_C)$. Equation (5-1) is a necessary requirement for consistency at a high enough temperature.

Now for model A at very high temperatures, i.e., very low "a" values, we know that we can approximate equation (3-3) by

$$G = G_0 [(1-e^{-x_0})/(1-e^{-a x_0})]$$

$$= G_0 [(1-e^{-x_0})/(a x_0)]$$

$$= G_0 [(1-e^{-x_0}) T/(T_0 x_0)] \qquad (x \ll 1) \qquad (5\text{-}2)$$

At inception of $G^{-1}$, $T_C$ replaces the temperature T. We can also substitute the values specified in equation (3-9). Inserting both into equation (5-2) allows us to write that at G formation,

$$G_C = .08455 \, G_0 \, T_C \qquad \text{(model A)} \qquad (5\text{-}3)$$

For model B, at very high temperatures, i.e., very small "a" values, equation (3-4) can also be approximated. In this instance,

$$G = G_0 [(\coth(x_0)-1/x_0)/(\coth(x)-1/x)]$$

$$= G_0 \, 3[(\coth(x_0)-1/x_0)/(x)]$$

$$= G_0 \, 3T \, [(\coth(x_0)-1/x_0)/ (T_0 \, x_0)] \quad (x \ll 1) \quad (5\text{-}4)$$

Again, G is proportional to T, just as in equation (5-2), for very small "x". In the second line of equation (5-4), use has been made of the fact that for small x, the Langevin function reduces to $L(x) \equiv \coth(x) - 1/x \cong x/3$. At inception of $G^{-1}$, TC replaces T. Furthermore, we can insert our values specified by equation (3-14). Substituting both into equation (5-4) gives in this situation

$$G_C = .0587 \, G_0 \, T_C \quad \text{(model B)} \quad (5\text{-}5)$$

We notice that for both models A and B, *$G_C$ is directly proportional to $T_C$*. This holds only in the limit of small x, or equivalently, small "a" values (very high temperatures). It must therefore hold true at $T = T_C$.

We next substitute equations (5-3) and (5-5) into our fundamental relation, equation (5-1). We start with equation (5-3). Substituting this into relation (5-1) gives

$$T_C \equiv (\hbar c^5/G_0 \, k_B{}^2)^{1/2} \, (.08455 \, T_C)^{-1/2}$$

$$= 1.42 * 10^{32} \, (.08455 \, T_C)^{-1/2} \quad (5\text{-}6)$$

Solving for $T_C$, we find

$$T_C = 6.20 * 10^{21} \, K \quad \text{(model A)} \quad (5\text{-}7)$$

For model B, we proceed analogously. We substitute equation (5-5) into equation (5-1). In this instance, we obtain

$$T_C \equiv (\hbar c^5/G_0 \, k_B{}^2)^{1/2} \, (.0587 \, T_C)^{-1/2}$$

$$= 1.42 * 10^{32} \, (.0587 \, T_C)^{-1/2} \quad (5\text{-}8)$$

Solving for $T_C$, we find

$$T_C = 7.01 * 10^{21} \, K \quad \text{(model B)} \quad (5\text{-}9)$$

The values obtained for models A and B are remarkably close to each other, not just order of magnitude wise. The temperatures are large, but still very well below the Planck temperature, which equals $1.42 * 10^{32}$ K. In fact, in retracing our steps, we find that we can approximate $T_C$ for both models A and B as $T_C \approx (T_P)^{2/3}$. $T_C$ is fundamental in our view whereas the Planck temperature, $T_P$, on the other hand, is not.

The associated scale for $G^{-1}$ formation is given next. For models A and B, using our temperatures (5-7) and (5-9), we find that

$$a_C = T_0/T_C = 4.37 * 10^{-22} \quad \text{(model A)} \quad (5\text{-}10a)$$

$$a_C = T_0/T_C = 3.89 * 10^{-22} \quad \text{(model B)} \quad (5\text{-}10b)$$

Because the $T_C$ values are very close in both models, it comes as no surprise that the "$a_C$" values are very close as well. The redshift at G formation is very high if we accept the scale factor determinations specified in equations (5-10).

In table V in appendix E, we have specified the scale parameters indicated by equations (5-10a) and (5-10b). We have also calculated accordingly the $G/G_0$ values at these scales, as well as other quantities, given our formulae above. For the two models presented, the values calculated are

$$G/G_0 = 5.273 * 10^{20} \qquad \text{(model A)} \qquad (5\text{-}11a)$$

$$G/G_0 = 4.113 * 10^{20} \qquad \text{(model B)} \qquad (5\text{-}11b)$$

Both values are similar and quite large. We believe, nevertheless, that there could very well be a twenty order of magnitude increase in G at these extremely high temperatures. At "a" = .001 which is close to recombination, G is still comparatively low in value, and that already puts us at a look-back time of 13.8 Gyr. For model A, $G/G_0$ equals 231, whereas for model B, $G/G_0$ = 160, at "a" = .001. See table I in appendix A. However, at these much higher temperatures of $10^{21}$ K, $G/G_0$ could be much larger. Again, this would go a long way towards explaining the vacuum energy discrepancy between present and past values.

The cosmological constant, $\Lambda$, is sometimes referred to as the mass of the vacuum. Due to our equation, (2-23), we can estimate its value at G formation. We substitute equations (5-11a) and (5-11b) into relations (2-23). This gives

$$\Lambda/\Lambda_0 = 2.78 * 10^{41} \qquad \text{(model A)} \qquad (5\text{-}12a)$$

$$\Lambda/\Lambda_0 = 1.69 * 10^{41} \qquad \text{(model B)} \qquad (5\text{-}12b)$$

We do not have the 122 order of magnitude difference between present and past values because *we are stopping well short of the Planck scale*. We have instead a 41 order of magnitude increase as indicated by equations (5-12a) and (5-12b). Dark energy does scale in our estimation, but never by nearly as much as either matter or radiation. The most radical scaling of dark energy occurs in the very early universe as it is there that the "w" values deviate significantly from negative one.

It has been argued that gravity may not be a fundamental force. Instead it may a low energy phenomenological limit which vanishes at incredibly high temperatures/energies, at the scales indicated by equations (5-7) and (5-9). If this were the case, then it would be interesting to calculate the radiative energy density, the dominant form of energy at these very temperatures. Using Planck's formula, $u = 4\sigma T^4/c$ where $\sigma$ = Stefan-Boltzmann constant = $5.67 * 10^{-8}$ J/(s m$^2$ K$^4$), we find that for model A we have an energy density of the order u ($T_C = 6.20 * 10^{21}$ K) = $1.12 * 10^{72}$ J/m$^3$. For model B, we have correspondingly, u ($T_C = 7.01 * 10^{21}$ K) = $1.83 * 10^{72}$ J/m$^3$. While high, they are nowhere near as large as those which one would obtain at a Planck temperature of $T_P = 1.42 * 10^{32}$ K, where in that instance, u = $3.07 * 10^{113}$ J/m$^3$. For radiation, the pressure is always one-third the energy density. Therefore, at these temperatures the radiative pressure is also quite large, but apparently not large enough to prevent G formation from occurring, if our picture is correct. Thermal quantum fluctuations in the vacuum have damped down sufficiently such that now a long-range correlation can establish itself *within the vacuum*. This is the way we imagine the onset of gravity. The scalar field of Jordan emerges at exactly this scale.

## VI  *SUMMARY AND CONCLUSIONS*

In order to provide a possible explanation for the cosmological constant fine tuning problem, we identified the quintessence parameter, w, with a time-varying gravitational constant, G. Specifically, $G/G_0 = a^{-\alpha}$, where "a" is the cosmic scale factor, $\alpha \equiv 3(1+w)$, and $w \equiv p_\Lambda/(\rho_\Lambda c^2)$. The quantities, $p_\Lambda$ and $\rho_\Lambda$, are the vacuum pressure and density, respectively. See equation (2-4). Using a current best value estimate for w, $w_0 = -.98$, we were able to show that that $\dot{G}/G|_0 = -.06 H_0$ where $H_0$ is the current value for Hubble's parameter. This result is in line with the original thinking of Dirac and Jordan, who advocated the view that G is of cosmological origin, is decreasing with respect to cosmological time, and is currently varying very slowly. Jordan, in particular, extended Dirac's idea and claimed that $\dot{G}/G = -H$, where H is Hubble's parameter. Our calculated result is in line with what Jordan claimed, but within the observational bounds placed on G, which currently requires that $\dot{G}/G < .1 H$.

We presented two specific parametrizations for G(a) where "a" is the cosmic scale parameter, $a = (1+z)^{-1} = T_0/T$, z is the redshift and T is the CMB temperature. Both parametrizations gave us remarkably similar results for $G/G_0$, look-back times, vacuum pressure, and onset of G formation. The first parametrization (model A) is based on a charging capacitor model where $G^{-1}$ saturates as T approaches zero. At high temperatures (energies), $G^{-1}$ has a low value. See equation (3-1). The second parametrization (model B) treats $G^{-1}$ as an order parameter, which acts much like the magnetization in a paramagnetic. $G^{-1}$ increases with decreasing T and at high temperatures, $G^{-1}$ is also very small. We have a Langevin function dependency in this instance. See equation (3-2). Both models are remarkably simple in that they depend only on background temperature (one-dimensional parametrizations), and even though physically and mathematically distinct, they both yield very similar results. Thus the tracking behavior for model A is practically identical to that of model B, i.e., $w(a)|_A \approx w(a)|_B$. With these parametrizations, we can give a specific evolution for G/G, $\dot{G}/G$, $\alpha$, w, etc. and these are given in table and graphical form. See table I in appendix A, and the graphs in appendix B, which illustrate some of these dependencies. See also appendix C, and the graphs contained therein, which is an equivalent formulation. In the $\Lambda$CDM model, G does not change, $w = -1$ and $\alpha$ is uniquely zero. Upon a comparison of models, it is only in the very early universe where drastic deviations from $\Lambda$CDM occur. The similarity in the results between models A and B leads us to suspect that there may be a universality of sorts behind the order parameter approach.

Using our model A and B parametrizations, we are, in the present epoch, in a phase where G is approximately constant. In fact, we are close to a saturated value for $G^{-1}$, which we call $G_\infty^{-1}$. For model A, we have calculated that $(G_\infty^{-1})|_A = 1.014 G_0^{-1}$. See equation (3-28). For model B, the calculation leads to $(G_\infty^{-1})|_B = 1.054 G_0^{-1}$. See equation (3-29). G will approach the saturated value in model A within a relatively short time, when the cosmic scale parameter has achieved a value "a" ≈ 2. For model B, the universe has to increase its size more dramatically, to an "a" value equal to "a" ≈ 10, or 10 times its current size. See graphs 4a and 4b in appendix B where this is illustrated.

In section IV, we considered the time evolution for a universe where $\dot{G}/G \neq 0$, specifically for our two models, A and B. Because the universe now evolves differently, we have compared our models A and B with the $\Lambda$CDM result. Model B is more conservative than model A in that it seems to track the $\Lambda$CDM results better. Even though the predicted age of the universe for our time-varying models are less, they are close. In fact, close enough, such that with minor revisions in input parameters, we can achieve a perfect match in predicted age with the concordance model. We argue that if the age of the universe is to be held constant at 13.8 Gyr, and if the age correction factor is to remain as it is in the concordance model, then the Hubble parameter has to be decreased somewhat in value. We give reasons why assuming a $H_0$ value close to 62.3 km/(s Mpc) may present an obvious solution to the problem of matching ages. We give results and graphs for non-adjusted and adjusted $H_0$ values. These are presented in table III and table IV in appendix D, as well as in graphs 6a and 6b. Another possible solution is to increase the dark energy contribution and lower the value of the matter density parameter. This would change the age correction factor, and bring it to a value, which makes the age of the universe line up with the predictions of the concordance model. In that scenario, $H_0$ would not have to be adjusted.

Finally, we have considered in section V, the inception of $G^{-1}$. Being an order parameter, it arose once long range order could be established against a violent background of disruptive high temperature vacuum fluctuations. Those fluctuations are due to virtual particle creation and annihilation. We determined that for both models A and B, the scale for $G^{-1}$ onset occurred at $a_C$ (A) = $4.37*10^{-22}$ for model A, and at $a_C$ (B) = $3.89*10^{-22}$ for model B, where "$a_C$" is the cosmic scale parameter at emergence of $G^{-1}$. These correspond to temperatures of $T_C/T_0 = 2.29*10^{21}$ K for model A, and $T_C/T_0 = 2.57*10^{21}$ K, for model B. The value for $G/G_0$ at these temperatures were found to be $G_C(A)/G_0 = 5.27*10^{20}$ for parametrization A, and $G_C(B)/G_0 = 4.12*10^{20}$ for parametrization B. The reader will note that these values are remarkably close to one another numerically, and furthermore, that the temperatures for $G^{-1}$ onset are well below the Planck temperature of $1.42*10^{32}$ K. In general, we have found within our models that $T_C \approx (T_P)^{2/3}$ is a good approximation for gravity formation. Therefore, gravity, as exemplified by the constant G, did not exist at the onset of the Big Bang. We believe that it came into being "much later", well past the inflation phase. If this is the case, then during inflation, the universe would not have been constrained by gravity *and there would have been no hindering force to prevent exponential expansion.*

With these values, we are in a position to explain, or at least dramatically alleviate, the cosmological vacuum fine tuning problem, accepting the notion that the present observed cosmological constant is related to the quantum vacuum. Using equation (2-23), we determine that at G inception, $\Lambda_{VACUUM} = 10^{41} \Lambda_{OBS}$. The $\Lambda_{VACUUM}$ is decreasing as $G^{-1}$ increases, by equation (2-23). This could make sense because in the very early universe, when T was very high, no long-range correlation could form. As quantum thermal fluctuations decreased, $G^{-1}$ could establish a foothold. The vacuum cosmological constant has decreased to its present low energy value, the value we observe and measure today. The key is to recognize the role of $G^{-1}$ in this evolution. Without a varying G, it would be difficult to imagine how the mass of the vacuum could change its value with respect to cosmological time.

Future work needs to be done before models of this nature can be accepted. We need to give a physical basis for our parametrizations A and B. We could entertain other parametrizations as other tracker solutions may lead to more interesting consequences, or have features that our naïve parametrizations are lacking. We could attempt to measure observationally both w and G to greater precision. As far as $\dot{G}/G$ is concerned, we believe we may be within striking range of a non-zero result, if equation (2-11) is to be believed. Both "w" and $\dot{G}/G$ need to be determined more accurately as this would ultimately decide whether they are related or not. One can consider the ramifications for the very early universe if G is non-existent before a certain point in time. What does this mean for inflation if gravity is switched off at temperature higher than $7*10^{21}$ K? What does this mean for the other interactions and for the GUT scale? What preceded gravity? How does a time varying G influence Jeans gravitational clumping, early star formation, galaxy formation, etc.? Would the evolution and structure of the universe change dramatically from the one we currently observe if G varies? What would this mean for black hole formation and the Schwarzschild radius in particular? These are all questions of considerable value and interest. Remember that G does not really increase dramatically within our models, unless "a" falls below .01. This corresponds to only 14 Myr after the Big Bang, significantly before early star formation, and galaxy evolution.

We conclude with a few remarks concerning a long-standing problem in physics, namely the renormalization of gravity. The dimensional character of Newton's constant has long been recognized as the single most important obstacle towards achieving a renormalizable, i.e., a quantum theory of gravity. If we have a coupling constant with an innate dimensionality, the quantum action changes to each order in perturbation theory. We thus have an infinite number of terms within the Lagrangian, which leads to a divergent series. The situation is not dissimilar to the weak interaction before it was combined into a renormalizable electro-weak interaction. There the Fermi constant, $G_F$, also has an inherent dimensionality, which incidentally is exactly of the same dimension as Newton's constant. In natural units, $\dim[G_F] = \dim[G_N] = (Mass)^{-2} = \dim^{-1}[Magnetization]$. By spontaneously breaking the electro-weak interaction, it was possible to give $G_F$ a mass. In fact, in the limit of low momentum/energy exchanges, the Fermi constant was shown to be inversely proportional to the mass of the $W^{+/-}$ boson squared.

We believe that we may have an analogous situation here with gravity. First, at very high energies, we argued that gravity might not exist. It is a low energy phenomenological limit, and only comes into being once a specific symmetry has been broken. Our order parameter, $G^{-1}$, is nothing else but the vacuum expectation value of a scalar field, $<0|\varphi^2|0>$. This is *not* to be identified with the quintessence field, as the quintessence field is defined differently. It can however be identified with the scalar field as originally defined by Jordan. At high energies, the VEV, $<0|\varphi^2|0>$, disappears. At low energies, it assumes a vacuum expectation value, the value we observe presently, which, incidentally, is very close to its saturation value. This does not vanish. The cosmic scales over which this happens, from onset of $G^{-1}$, to near saturation value, $G_\infty^{-1}$, is called the coherence length. What makes this so spectacular is the very large range involved. This coherence length spans a cosmic scale range in excess of 23 orders of magnitude. The cosmic scale factor at $G^{-1}$ formation was about $10^{-22}$ and it will continue to about 10 before saturation is reached! During this time, the value of $G^{-1}$ increases from zero to effectively $G_\infty^{-1}$. Recognizing that $G_\infty^{-1}$ is very close to $G_0^{-1}$, we end up with a very large mass squared as a result of spontaneous symmetry breaking, but only in the present epoch. In a much earlier epoch, the mass being proportional to $M^2 \sim <0|\varphi^2|0> \sim G^{-1}$ was much, much smaller.

The graphs 5a and 5b in appendix C give the evolution of the order parameter, with its first derivative. As such it applies to $<0|\varphi^2|0>$, the (mass of the vacuum)$^2$. In closing, we can interpret $G^{-1}$ as the VEV, $<0|\varphi^2|0>$, and $G_0^{-1}$ with another VEV, $<0|\varphi_0^2|0>$, the present day value. In addition, the ratio, $G^{-1} / G_0^{-1}$ can be interpreted as a ratio of vacuum expectation values, $<0|\varphi^2|0>/<0|\varphi_0^2|0>$. By analogy to magnetization, we can call $G^{-1} = <0|\varphi^2|0>$, the "gravitization" of the vacuum. Its value is close to being 100% achieved at present, but we believe that it may once have had a value, which was very, very much smaller.

The author would like to thank Gonzaga University and the physics department in particular, for their support and encouragement. Special thanks goes to Professors Eric Aver, Adam Fritsch, and Matt Geske, for reading the manuscript, and offering helpful comments and suggestions. Any shortcomings, however, are entirely those of author.

**REFERENCES:**

[1] S. Weinberg, Rev. Mod. Phys. 61, 1 (1989)

[2] Copeland, E. J., Sami, M., & Tsujikawa, S., Dynamics of Dark Energy. 2006, International Journal of Modern Physics D, 15, 1753, arXiv:hep-th/0603057

[3] Tsujikawa, S., Modified gravity models of dark energy. 2010, Lect. Notes Phys., 800, 99, arXiv:1101.0191

[4] Y. Fujii, Phys. Rev. D 26, 2580 (1982); L. H. Ford, Phys. Rev. D 35, 2339 (1987)

[5] C. Wetterich, Nucl. Phys B. 302, 668 (1988)

[6] T. Chiba, N. Sugiyama and T. Nakamura, Mon. Not. Roy. Astron. Soc. 289, L5 (1997);

[7] P. G. Ferreira and M. Joyce, Phys. Rev. Lett. 79, 4740 (1997)

[8] R. R. Caldwell, R. Dave and P. J. Steinhardt, Phys. Rev. Lett. 80, 1582 (1998)

[9] P. J. E. Peebles and B. Ratra, Astrophys. J. 325, L17 (1988)

[10] B. Ratra and J. Peebles, Phys. Rev D 37, 321 (1988)

[11] Komatsu E., et al. [WMAP Collaboration], Astrophys. J. Suppl. **192** (2011) 18 [arXiv: 0803.0547 [astro-ph]]. Current best estimate for w is $w = -.98 +/- .053$ = value as quoted in text

[12] Lahav, O and Liddle, A.R., Cosmological Parameters, Nov. 2015, value quoted is $w = -.97 +/- .05$ compilation of CMB, SN and BAO measurements assuming a flat universe

[13] Collaboration, Planck, PAR Ade, N Aghanim, C Armitage-Caplan, M Arnaud, et al., Planck 2015 results. XIII. Cosmological parameters. arXiv preprint 1502.1589v2[1 ] (https://arXiv.org/abs/1502.1589v2), 6 Feb 2015

[14] Planck Collaboration XIV, Planck 2015 results. XIV. Dark energy and modified gravity. 2016, A&A, in press, arXiv: 1502.01590. The value quoted is $w = -1.006 +/- .045$ but this is not the most secure estimate; the tightest constraint is obtained for the Planck TT + low P

+ BSH combination. This gives w > -1, and is centered about w = - .98. See figure 28 (section 6.3) in the former publication and figures 3a and 4a (section 5.1), in the latter publication.

[15] See reference [11], op. cit.

[16] See reference [13], op. cit.

[17] See reference [11], op. cit.

[18] See reference [13], op. cit. The value quoted is $\Omega_k = -.005^{+.016}_{-.017}$ (95% CL, Planck TT + low P + lensing). See section 6.2.4 on curvature, in particular, equation (49).

[19] P. A. M. Dirac (1937). "The Cosmological Constants", *Nature*. **139** (3512): 323. Bibcode: 1937 Natur.139. 323D.doi:10.1038/139323a0.

[20] P. A. M. Dirac (1938). "A New Basis for Cosmology" *Proceedings of the Royal Society of London A*. **165** (921): 199–208. Bibcode: 1938RSPSA. 165. 199D. doi:10.1098/rspa.1938.0053.

[21] P. A. M. Dirac (1974). "Cosmological Models and the Large Numbers Hypothesis "*Proceedings of the Royal Society of London A*. **338** (1615): 439–446. Bibcode: 1974 RSPSA.338.439D. doi:10.1098/rspa.1974.0095.

[22] P. Jordan, "G has to be a field", Naturwiss. 25, 513 (1937)

[23] P. Jordan "Formation of Stars and Development of the Universe" Nature, 164 (1949), pp. 637-640

[24] Die Wissenschaft, vol. 124. Friedrich Vieweg & Sohn, Braunschweig 1966 Heinz Haber: "Die Expansion der Erde" [The expansion of the Earth] *Unser blauer Planet* [*Our blue planet*]. Rororo Sachbuch [Rororo nonfiction] (in German) (Rororo Taschenbuch Ausgabe [Rororo pocket edition] ed.). Reinbek: Rowohlt Verlag. 1967 [1965]. pp. 48, 52, 54–55.

[25] Kragh, Helge (2015). "Pascual Jordan, Varying Gravity, and the Expanding Earth". *Physics in Perspective*. **17** (2):107. Bibcode: 2015PhP...17..107K (http://adsabs.harvard.edu/abs/2015PhP...17..107K). doi: 10.1007/s00016-015-0157-9 (https://doi.org/10.1007%2Fs00016-015-0157-9.)

[26] See reference [13], op. cit.

[27] See reference [24], op. cit.

[28] See reference [25], op. cit.

[29] Muller, P. M., 1978, in *On the Measurement of Cosmological Variations of the Gravitational Constant*, edited by L. Halphern (University of Florida, Gainesville, FL), p. 93.

[30] I.I. Shapiro, W.B. Smith, and M.B. Ash, Gravitational Constant: Experimental Bound on Its Time Variation, Phys. Rev. Lett. 26, 27 (1971). 1964-1969. They find that $|\dot{G}/G| < 4 *10^{-10}$ yr$^{-1}$.

[31] Van Flandern, T. C., 1971, Astron. J. **76**, 81.

[32] Van Flandern, T. C., 1975, Mon. Not. R. Astron. Soc. **170**, 333. $\dot{G}/G = 8 +/- 5 *10^{-11}$ yr$^{-1}$

[33] J. A. P. Martins, "Cosmology with varying constants," Philos. Trans. R. Soc. London, Ser. A **360**, 2681–2695 2002

[34] J. P. Uzan (2003). "The fundamental constants and their variation, Observational status and theoretical motivations". *Reviews of Modern Physics*. **75** (2): 403. arXiv:hep-ph/ 0205340 (https://arxiv.org/abs/hep-ph/0205340). Bibcode: 2003RvMP...75..403U (http://adsabs.harvard.edu/abs/2003RvMP...75..403U). doi:10.1103/RevModPhys.75.403 (https://doi.org/10.1103%2FRevModPhys.75.403.)

[35] J. D. Barrow, "Varying constants," Philos. Trans. R. Soc. London, Ser. A **363**, 2139–2153 _2005_

[36] J. P. Uzan (2011), Varying Constants, Gravitation and Cosmology, Living Rev. Relativ. 2011; 14(1): 2 doi: 10.12942/lrr-2011-2

[37] K. Nordtvedt, 1968, Phys. Rev. 170, 1186.

[38] K. Nordtvedt, 1988, Phys. Rev. D **37**, 1070.

[39] K. Nordtvedt, 1990, Phys. Rev. Lett. **65**, 953.

[40] S.E. Thorsett, The Gravitational Constant, The Chandrasekhar Limit, And Neutron Star Masses, Phys. Rev. Lett. 77, 1432 (1996). Here, $\dot{G}/G = - .6 +/- 4.2 * 10^{-12}$ yr$^{-1}$.

[41] E. Gaztañaga, E. García-Berro, J. Isern, E. Bravo, and I. Domínguez, "Bounds on the possible evolution of the gravitational constant from cosmological type-Ia supernovae," Phys. Rev. D **65**, 023506-1–9 _2001_.

[42] E. García-Berro, Y. Kubyshin, P. Loren-Aguilar, and J. Isern, "The variation of the gravitational constant inferred from the Hubble diagram of type Ia supernovae," Int. J. Mod. Phys. D **15**, 1163–1174 _2006_.

[43] García-Berro, J. Isern, and Y. A. Kubyshin, "Astronomical measurements and constraints on the variability of fundamental constants," Astron. Astrophys. Rev. **14**, 113–170 _2007_.

[44] See reference [35], op. cit.

[45] See reference [36], op. cit.

[46] See reference [13], op. cit.

[47] See, e.g., Wikipedia and look under "Age of Universe".

[48] See reference [13], op. cit.

[49] Freedman, W. L., Madore, B. F., Scowcroft, V., et al., Carnegie Hubble Program: A Mid-infrared Calibration of the Hubble Constant. 2012, ApJ, 758, 24, arXiv: 1208:3281

[50] Tammann, G. A. & Reindl, B., The luminosity of supernovae of type Ia from tip of the red-giant branch distances and the value of H0. 2013, A&A, 549, A136, arXiv:1208.5054. G. Tammann collaborated with Sandage, who was one of the first to determine a modern value for $H_0$.

**APPENDIX A**                       **TABLE I**

| "a" | $G/G_0$ (A) | $G/G_0$ (B) | $\dot{G}/G \vert_A$ | $\dot{G}/G \vert_B$ | α(A) | α(B) | w(A) | w(B) |
|---|---|---|---|---|---|---|---|---|
| 1    | 1.00E+00 | 1.00E+00 | -0.060 | -0.060 | 0.060 | 0.060 | -0.980 | -0.980 |
| 0.9  | 1.01E+00 | 1.01E+00 | -0.084 | -0.067 | 0.071 | 0.063 | -0.976 | -0.979 |
| 0.8  | 1.02E+00 | 1.02E+00 | -0.115 | -0.076 | 0.086 | 0.068 | -0.971 | -0.977 |
| 0.7  | 1.04E+00 | 1.03E+00 | -0.158 | -0.088 | 0.105 | 0.073 | -0.965 | -0.976 |
| 0.6  | 1.07E+00 | 1.04E+00 | -0.213 | -0.104 | 0.129 | 0.080 | -0.957 | -0.973 |
| 0.5  | 1.12E+00 | 1.06E+00 | -0.285 | -0.128 | 0.160 | 0.089 | -0.947 | -0.970 |
| 0.4  | 1.20E+00 | 1.10E+00 | -0.377 | -0.165 | 0.202 | 0.103 | -0.933 | -0.966 |
| 0.3  | 1.36E+00 | 1.16E+00 | -0.492 | -0.232 | 0.258 | 0.125 | -0.914 | -0.958 |
| 0.2  | 1.71E+00 | 1.31E+00 | -0.632 | -0.377 | 0.335 | 0.169 | -0.888 | -0.944 |
| 0.1  | 2.83E+00 | 1.91E+00 | -0.801 | -0.702 | 0.452 | 0.281 | -0.849 | -0.906 |
|      |          |          |        |        |       |       |        |        |
| 0.1  | 2.83E+00 | 1.91E+00 | -0.801 | -0.702 | 0.452 | 0.281 | -0.849 | -0.906 |
| 0.09 | 3.08E+00 | 2.06E+00 | -0.820 | -0.745 | 0.468 | 0.300 | -0.844 | -0.900 |
| 0.08 | 3.40E+00 | 2.25E+00 | -0.839 | -0.787 | 0.485 | 0.322 | -0.838 | -0.893 |
| 0.07 | 3.81E+00 | 2.51E+00 | -0.858 | -0.829 | 0.503 | 0.346 | -0.832 | -0.885 |
| 0.06 | 4.35E+00 | 2.86E+00 | -0.877 | -0.869 | 0.523 | 0.374 | -0.826 | -0.875 |
| 0.05 | 5.12E+00 | 3.37E+00 | -0.897 | -0.905 | 0.545 | 0.405 | -0.818 | -0.865 |
| 0.04 | 6.27E+00 | 4.14E+00 | -0.917 | -0.937 | 0.570 | 0.441 | -0.810 | -0.853 |
| 0.03 | 8.18E+00 | 5.44E+00 | -0.937 | -0.964 | 0.599 | 0.483 | -0.800 | -0.839 |
| 0.02 | 1.20E+01 | 8.07E+00 | -0.958 | -0.984 | 0.636 | 0.534 | -0.788 | -0.822 |
| 0.01 | 2.35E+01 | 1.61E+01 | -0.979 | -0.996 | 0.686 | 0.603 | -0.771 | -0.799 |
| **0.01**  | 2.35E+01 | 1.61E+01 | -0.979 | -0.996 | 0.686 | 0.603 | -0.771 | -0.799 |
| **0.009** | 2.61E+01 | 1.78E+01 | -0.981 | -0.997 | 0.692 | 0.612 | -0.769 | -0.796 |
| **0.008** | 2.93E+01 | 2.00E+01 | -0.983 | -0.997 | 0.700 | 0.621 | -0.767 | -0.793 |
| **0.007** | 3.34E+01 | 2.29E+01 | -0.985 | -0.998 | 0.707 | 0.631 | -0.764 | -0.790 |
| **0.006** | 3.89E+01 | 2.67E+01 | -0.987 | -0.999 | 0.716 | 0.642 | -0.761 | -0.786 |
| **0.005** | 4.66E+01 | 3.21E+01 | -0.989 | -0.999 | 0.725 | 0.654 | -0.758 | -0.782 |
| **0.004** | 5.81E+01 | 4.01E+01 | -0.991 | -0.999 | 0.736 | 0.668 | -0.755 | -0.777 |
| **0.003** | 7.73E+01 | 5.34E+01 | -0.994 | -1.000 | 0.748 | 0.685 | -0.751 | -0.772 |
| **0.002** | 1.16E+02 | 8.01E+01 | -0.996 | -1.000 | 0.764 | 0.705 | -0.745 | -0.765 |
| **0.001** | 2.31E+02 | 1.60E+02 | -0.998 | -1.000 | 0.788 | 0.735 | -0.737 | -0.755 |
| | | | | | | | | |
| **1**  | 1.00E+00 | 1.00E+00 | -0.060 | -0.060 | 0.060 | 0.060 | -0.980 | -0.980 |
| **2**  | 9.86E-01 | 9.71E-01 | -0.002 | -0.029 | 0.020 | 0.043 | -0.993 | -0.986 |
| **3**  | 9.86E-01 | 9.62E-01 | 0.000  | -0.019 | 0.013 | 0.036 | -0.996 | -0.988 |
| **4**  | 9.86E-01 | 9.57E-01 | 0.000  | -0.014 | 0.010 | 0.032 | -0.997 | -0.989 |
| **5**  | 9.86E-01 | 9.54E-01 | 0.000  | -0.011 | 0.009 | 0.029 | -0.997 | -0.990 |
| **6**  | 9.86E-01 | 9.52E-01 | 0.000  | -0.010 | 0.008 | 0.027 | -0.997 | -0.991 |
| **7**  | 9.86E-01 | 9.51E-01 | 0.000  | -0.008 | 0.007 | 0.026 | -0.998 | -0.991 |
| **8**  | 9.86E-01 | 9.50E-01 | 0.000  | -0.007 | 0.007 | 0.025 | -0.998 | -0.992 |
| **9**  | 9.86E-01 | 9.49E-01 | 0.000  | -0.006 | 0.006 | 0.024 | -0.998 | -0.992 |
| **10** | 9.86E-01 | 9.49E-01 | 0.000  | -0.006 | 0.006 | 0.023 | -0.998 | -0.992 |

# APPENDIX B

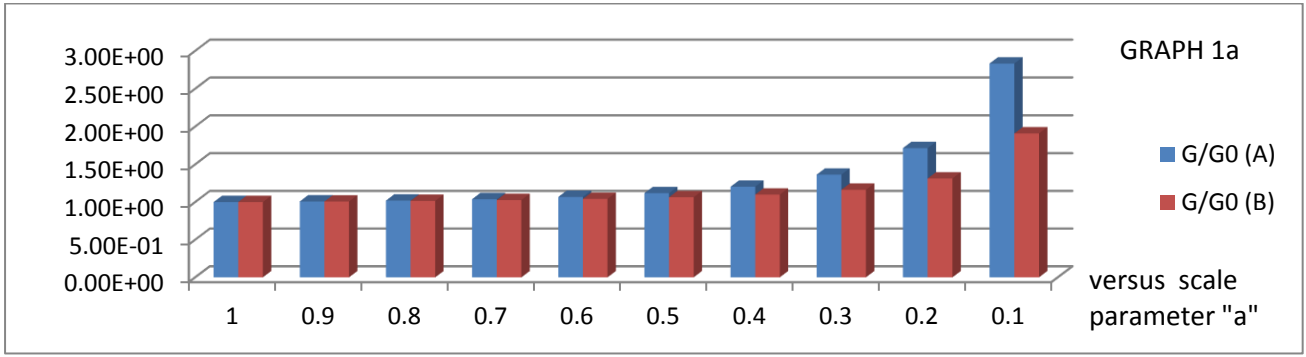

GRAPH 1a: G/G0 (A), G/G0 (B) versus scale parameter "a"

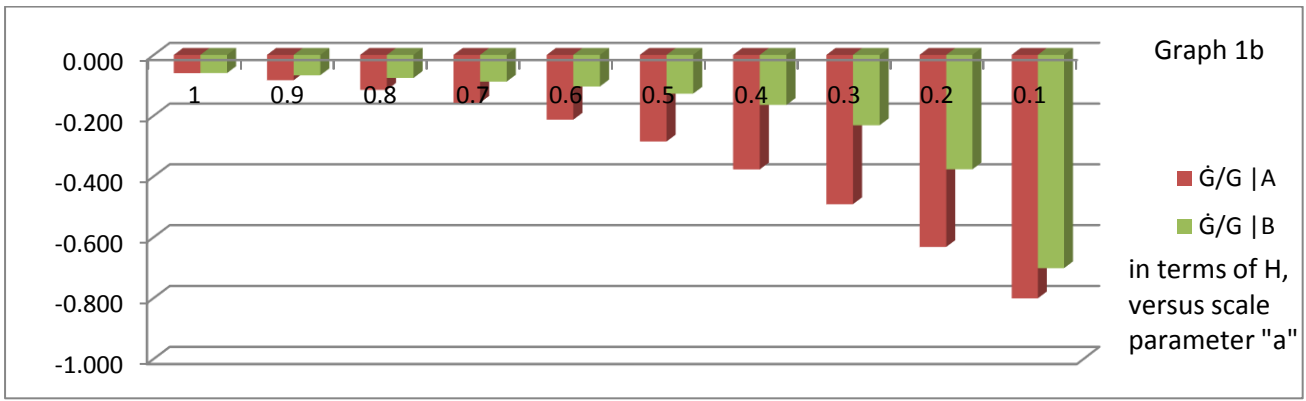

Graph 1b: Ġ/G |A, Ġ/G |B in terms of H, versus scale parameter "a"

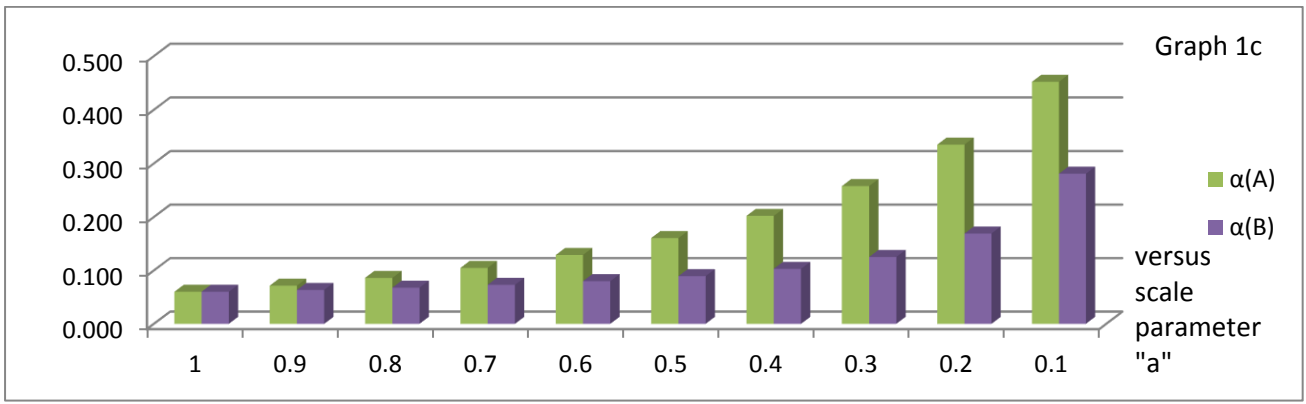

Graph 1c: α(A), α(B) versus scale parameter "a"

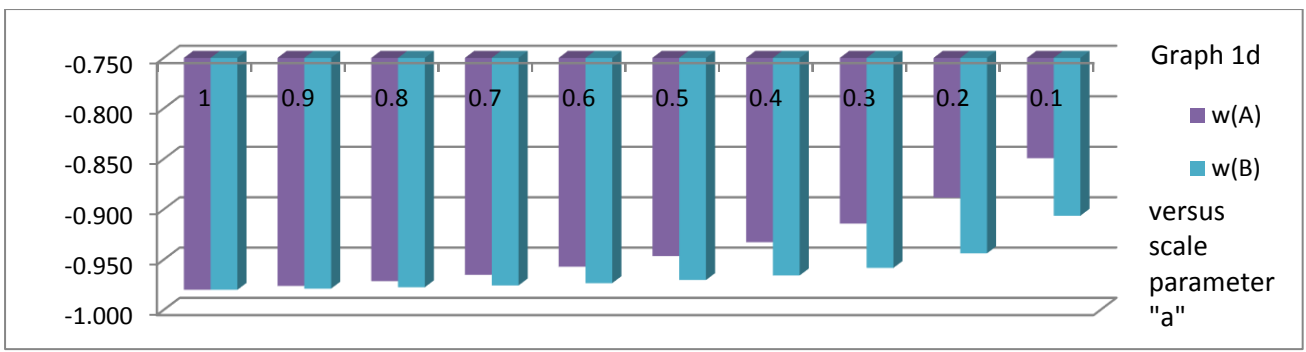

Graph 1d: w(A), w(B) versus scale parameter "a"

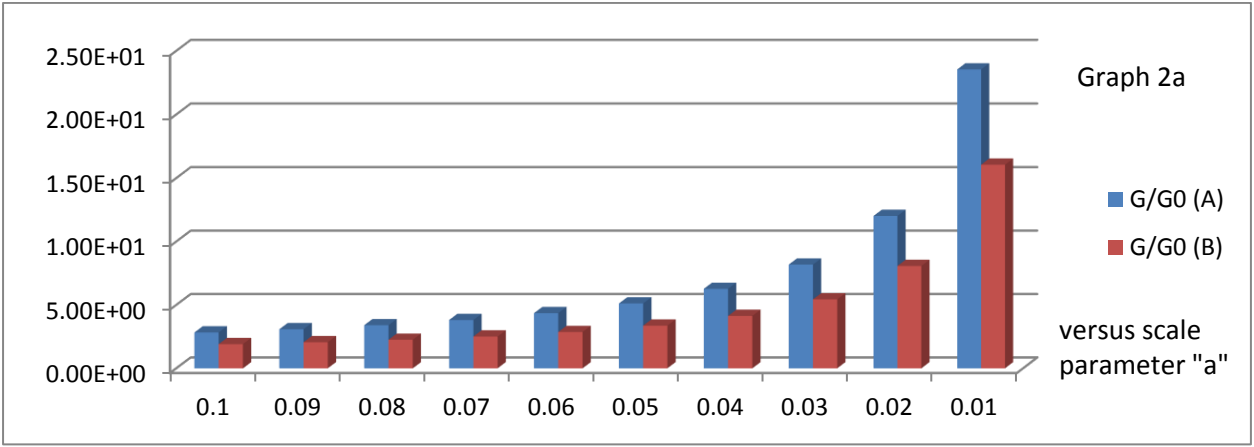
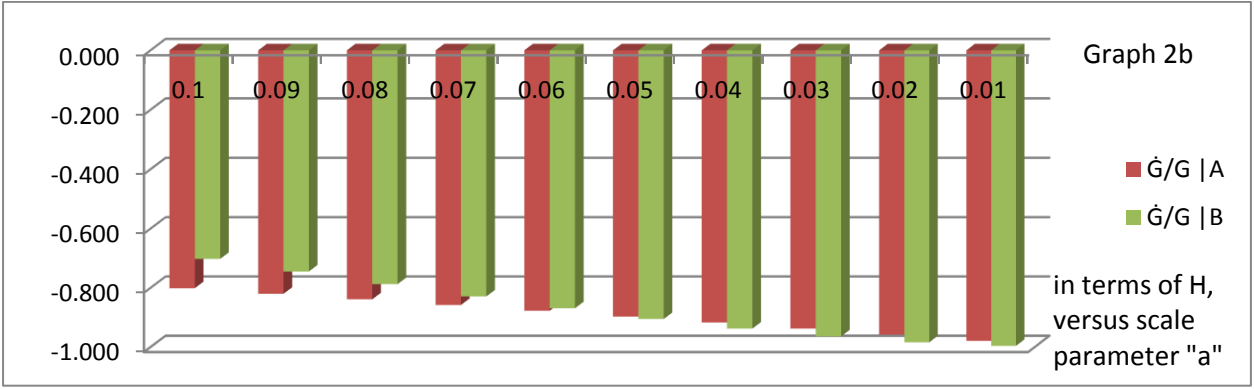
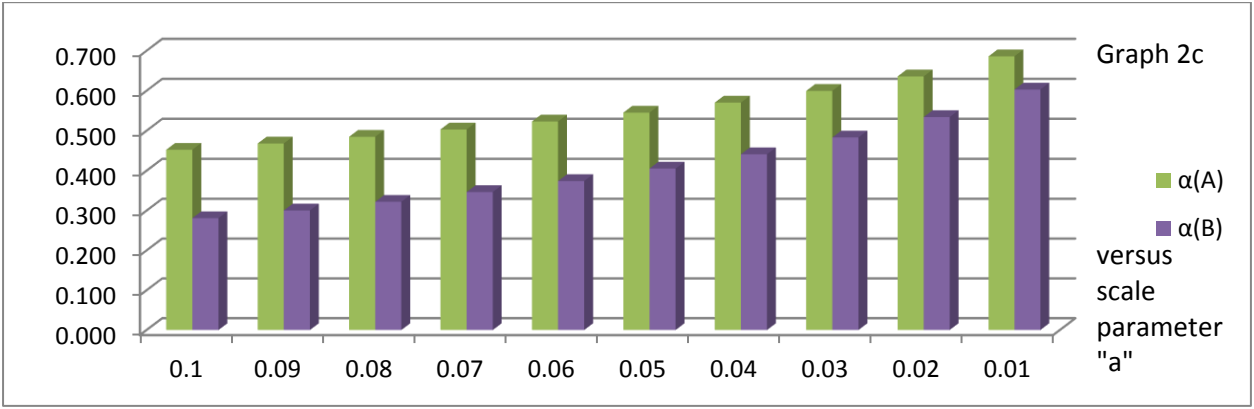
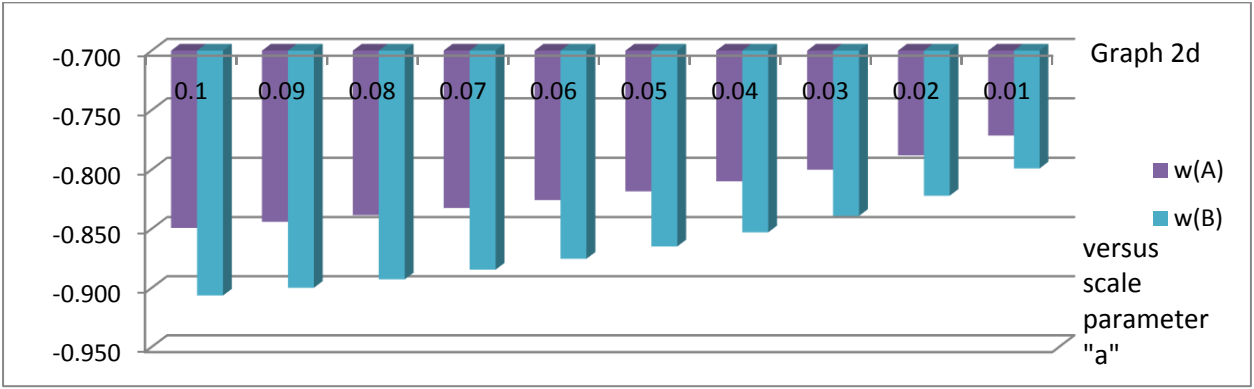

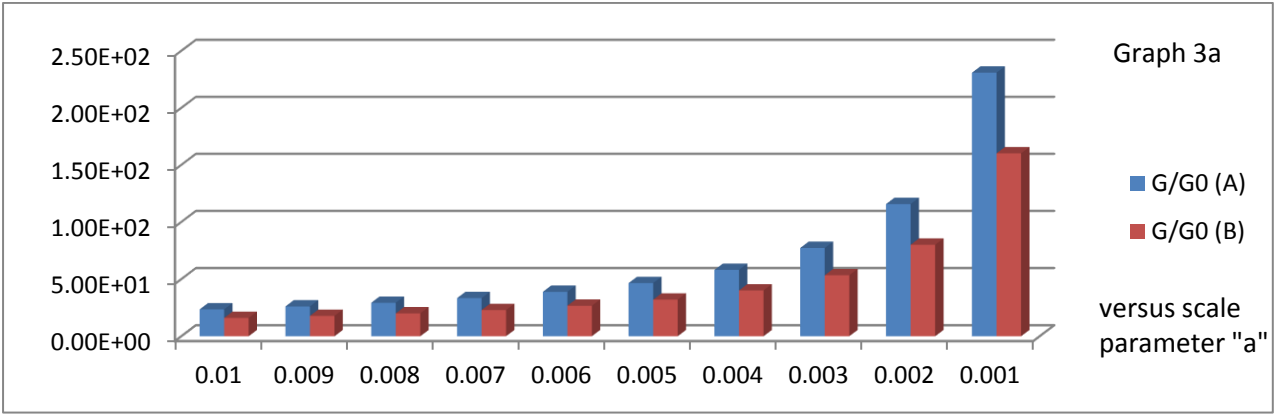
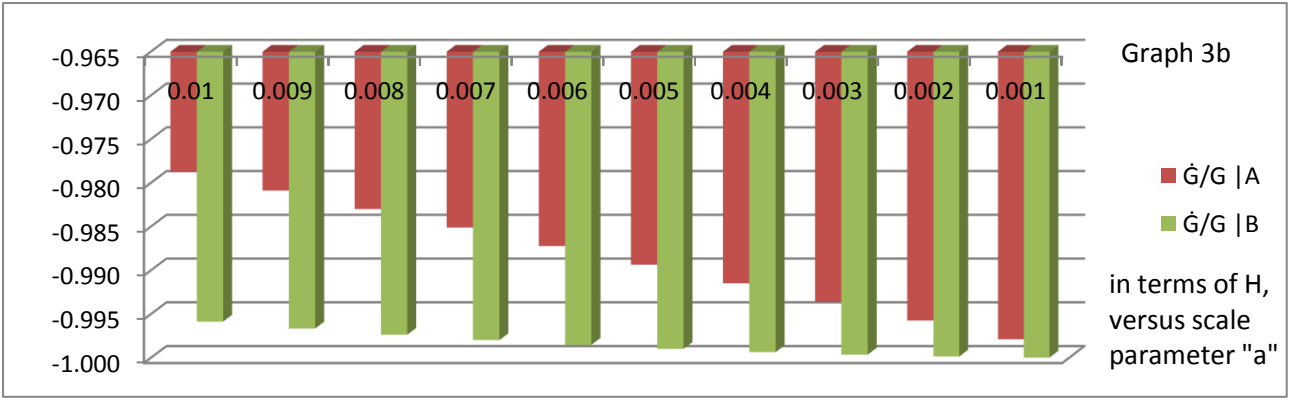
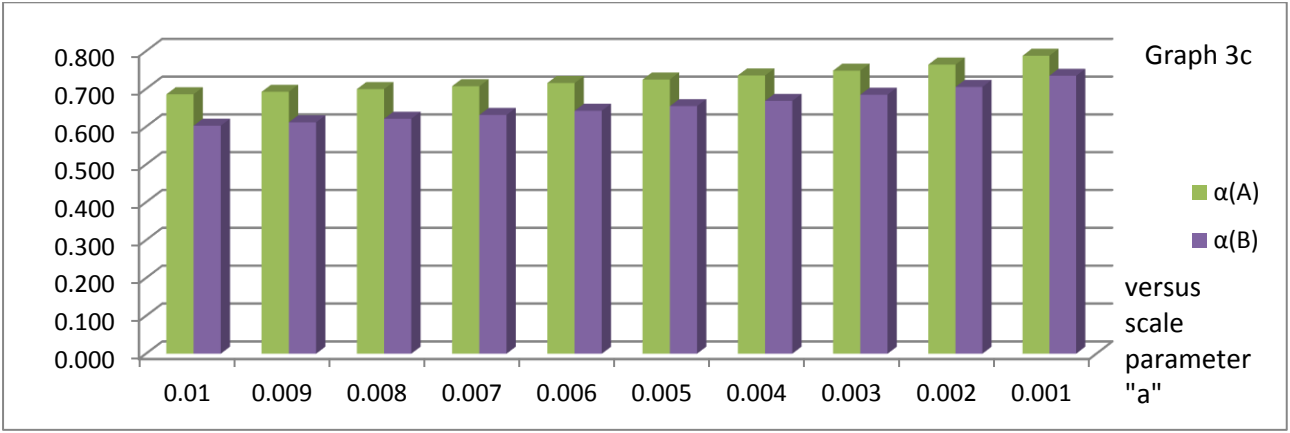
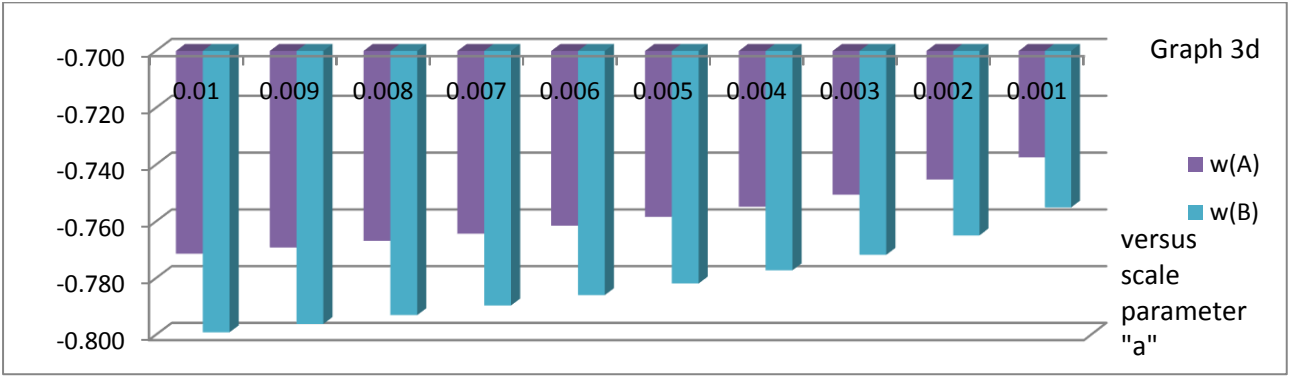

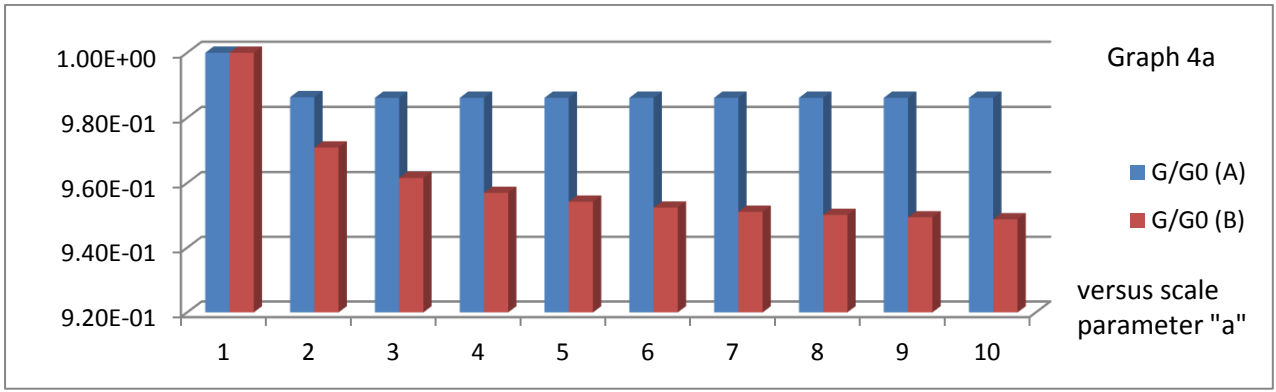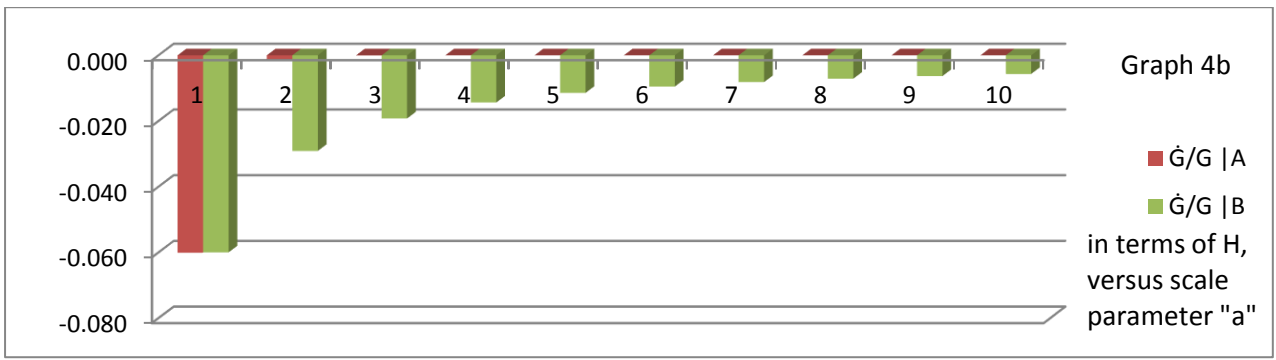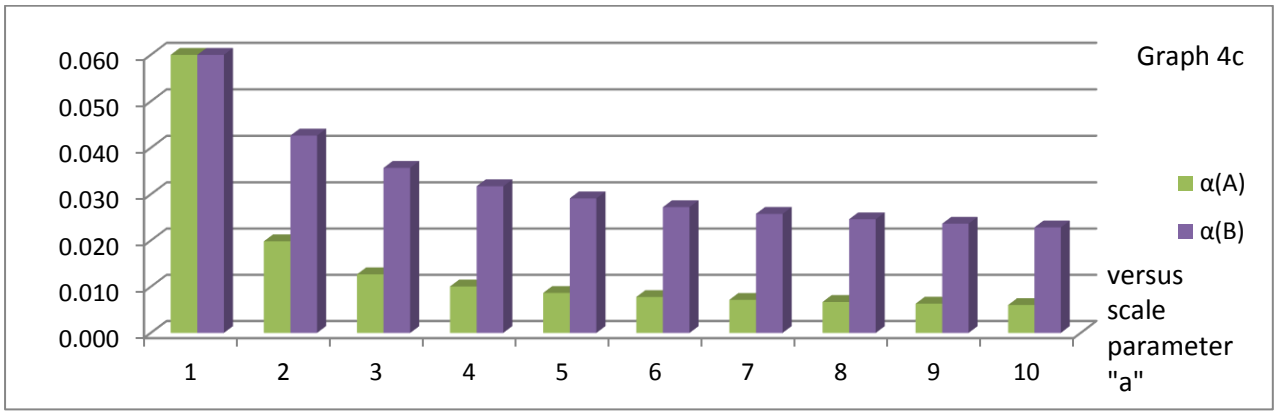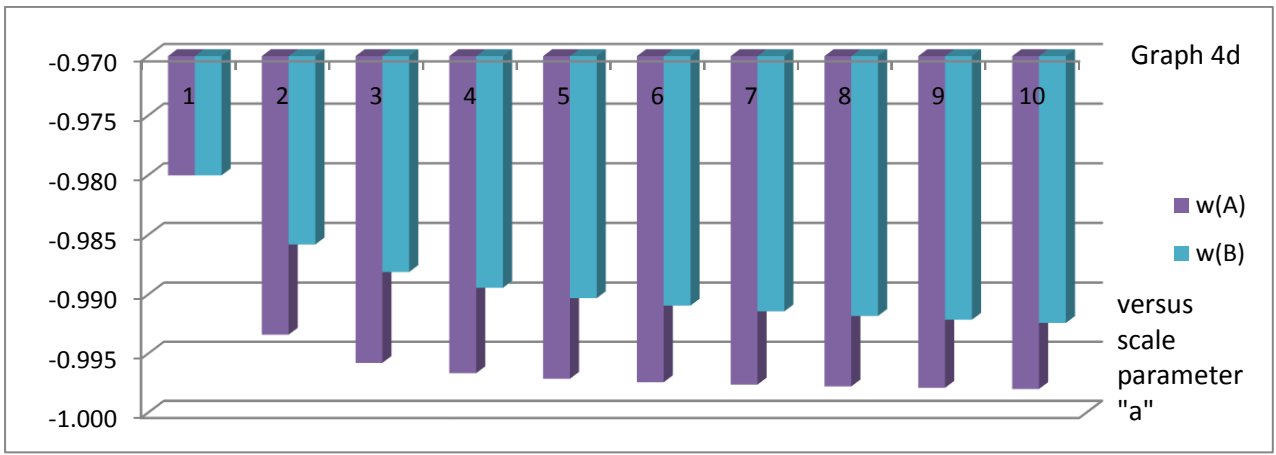

**APPENDIX C**                          **TABLE II**

| scale "a" | $G^{-1}/G_0^{-1}\vert_A$ | $G^{-1}/G_0^{-1}\vert_B$ | $\{d(G^{-1})/dt\}/G^{-1}\vert_A$ | $\{d(G^{-1})/dt\}/G^{-1}\vert_B$ |
|---|---|---|---|---|
| **0.05** | 1.95E-01 | 2.97E-01 | 0.897 | 0.905 |
| **0.1**  | 3.53E-01 | 5.24E-01 | 0.801 | 0.702 |
| **0.15** | 4.80E-01 | 6.71E-01 | 0.713 | 0.512 |
| **0.2**  | 5.83E-01 | 7.62E-01 | 0.632 | 0.377 |
| **0.25** | 6.66E-01 | 8.20E-01 | 0.559 | 0.289 |
| **0.3**  | 7.33E-01 | 8.60E-01 | 0.492 | 0.232 |
| **0.35** | 7.87E-01 | 8.89E-01 | 0.431 | 0.193 |
| **0.4**  | 8.31E-01 | 9.10E-01 | 0.377 | 0.165 |
| **0.45** | 8.66E-01 | 9.27E-01 | 0.329 | 0.144 |
| **0.5**  | 8.95E-01 | 9.40E-01 | 0.285 | 0.128 |
| **0.55** | 9.18E-01 | 9.51E-01 | 0.247 | 0.115 |
| **0.6**  | 9.36E-01 | 9.60E-01 | 0.213 | 0.104 |
| **0.65** | 9.51E-01 | 9.68E-01 | 0.184 | 0.095 |
| **0.7**  | 9.63E-01 | 9.74E-01 | 0.158 | 0.088 |
| **0.75** | 9.73E-01 | 9.80E-01 | 0.135 | 0.082 |
| **0.8**  | 9.81E-01 | 9.85E-01 | 0.115 | 0.076 |
| **0.85** | 9.87E-01 | 9.89E-01 | 0.098 | 0.071 |
| **0.9**  | 9.93E-01 | 9.93E-01 | 0.084 | 0.067 |
| **0.95** | 9.97E-01 | 9.97E-01 | 0.071 | 0.063 |
| **1**    | 1.00E+00 | 1.00E+00 | 0.060 | 0.060 |

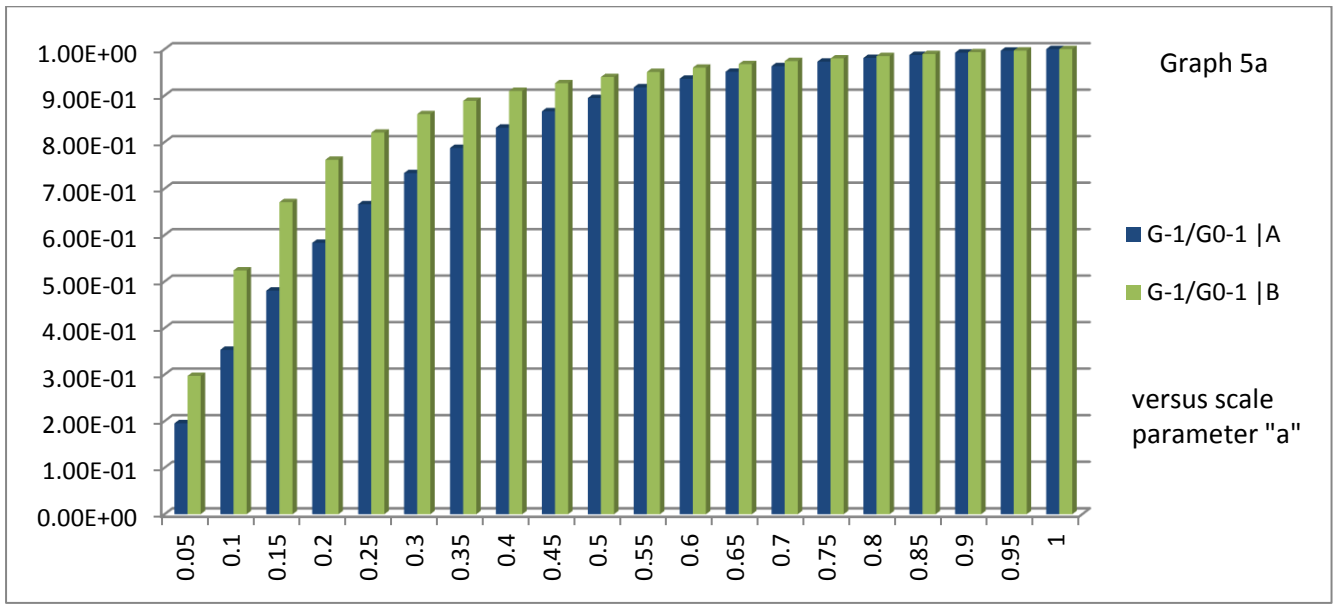

Graph 5a

G-1/G0-1 |A
G-1/G0-1 |B

versus scale parameter "a"

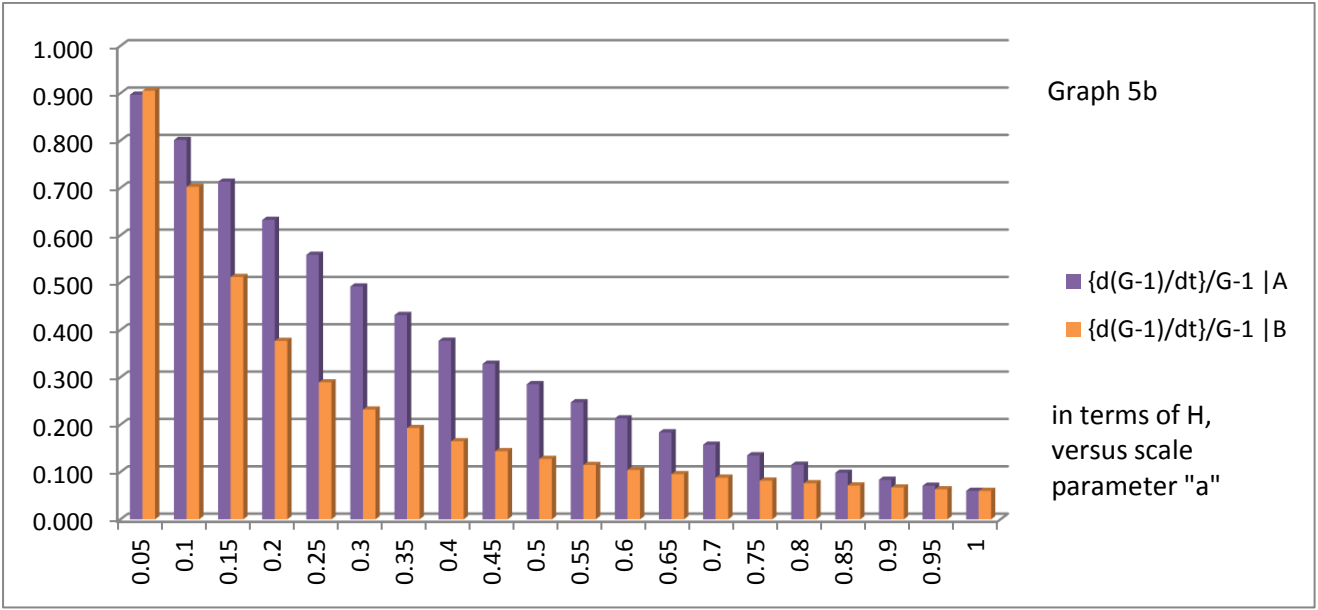

Graph 5b

in terms of H, versus scale parameter "a"

**APPENDIX D                TABLE III**

**LOOK BACK TIME in UNITS of $H_0^{-1}$**

| "a" | $(t_0 - t)$ in ΛCDM | $(t_0 - t)$ in Model A | $(t_0 - t)$ in Model B |
|---|---|---|---|
| 1 | 0 | 0 | 0 |
| 0.9 | 0.1026 | 0.1025 | 0.1025 |
| 0.8 | 0.2103 | 0.2094 | 0.2095 |
| 0.7 | 0.3219 | 0.3195 | 0.3201 |
| 0.6 | 0.4360 | 0.4307 | 0.4322 |
| 0.5 | 0.5497 | 0.5396 | 0.5431 |
| 0.4 | 0.6595 | 0.6418 | 0.6488 |
| 0.3 | 0.7609 | 0.7317 | 0.7444 |
| 0.2 | 0.8491 | 0.8036 | 0.8241 |
| 0.1 | 0.9181 | 0.8514 | 0.8804 |
| 0 | 0.9559 | 0.8689 | 0.9018 |

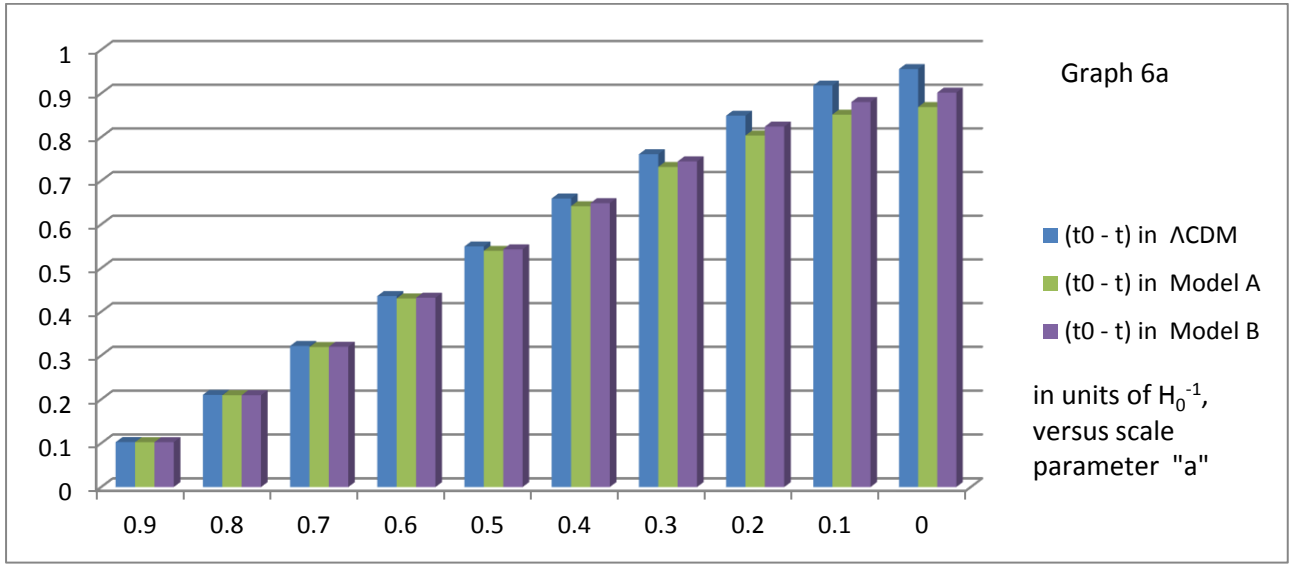

Graph 6a

$(t_0 - t)$ in ΛCDM
$(t_0 - t)$ in Model A
$(t_0 - t)$ in Model B

in units of $H_0^{-1}$, versus scale parameter "a"

## APPENDIX D          TABLE IV

**HUBBLE -ADJUSTED LOOK-BACK TIMES in UNITS of $H_0^{-1}$**

| "a" | $(t_0 - t)$ in ΛCDM Model | $(t_0 - t)$ in Model A | $(t_0 - t)$ in Model B |
|---|---|---|---|
| 1   | 0      | 0      | 0      |
| 0.9 | 0.1026 | 0.1128 | 0.1086 |
| 0.8 | 0.2103 | 0.2304 | 0.2221 |
| 0.7 | 0.3219 | 0.3515 | 0.3393 |
| 0.6 | 0.4360 | 0.4739 | 0.4581 |
| 0.5 | 0.5497 | 0.5937 | 0.5757 |
| 0.4 | 0.6595 | 0.7061 | 0.6877 |
| 0.3 | 0.7609 | 0.8051 | 0.7891 |
| 0.2 | 0.8491 | 0.8842 | 0.8735 |
| 0.1 | 0.9181 | 0.9368 | 0.9332 |
| 0   | 0.9559 | 0.9560 | 0.9559 |

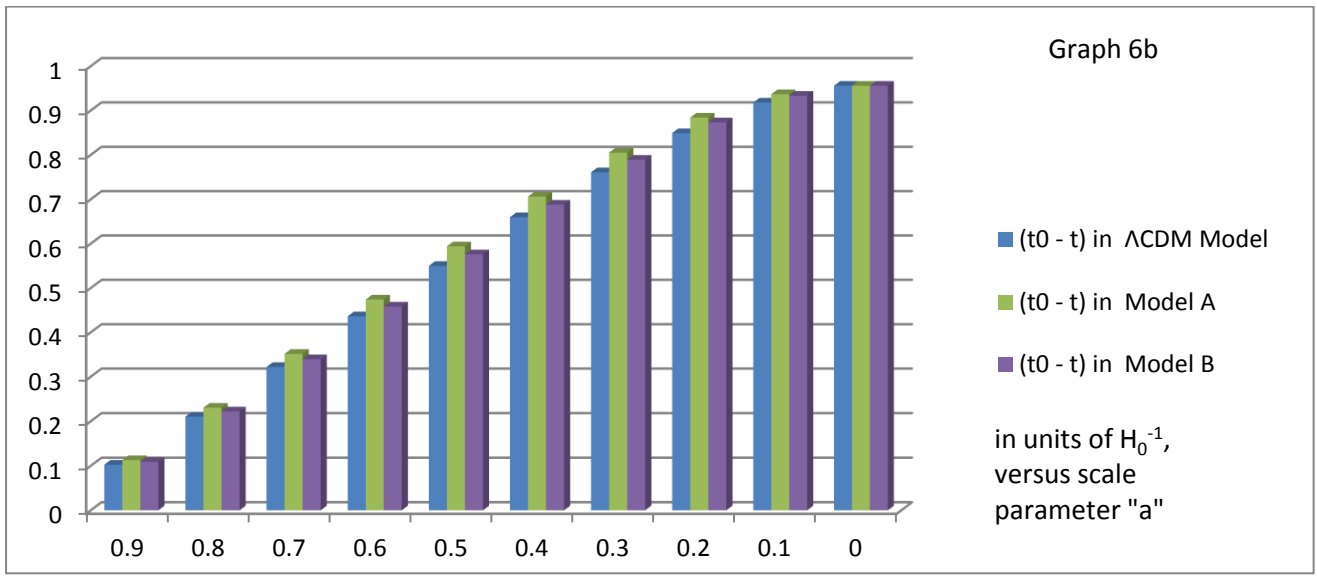

Graph 6b

$(t_0 - t)$ in ΛCDM Model, $(t_0 - t)$ in Model A, $(t_0 - t)$ in Model B, in units of $H_0^{-1}$, versus scale parameter "a"

## APPENDIX E          TABLE V

|         | $a_C$    | $T_C$ (K) | $G_C/G_0$ | $\dot{G}/G\,(a_C)$ | $\alpha\,(a_C)$ | $w\,(a_C)$ |
|---------|----------|-----------|-----------|--------------------|-----------------|------------|
| Model A | 4.37E-22 | 6.20E+21  | 5.27E+20  | (-1)(H)            | 0.970           | -0.677     |
| Model B | 3.89E-22 | 7.01E+21  | 4.12E+20  | (-1)(H)            | 0.963           | -0.679     |